%% file: sample62.tex
\documentclass{aastex63}

\usepackage{braket}
\usepackage{bm}
\usepackage{amsmath,amssymb}
\usepackage{subfigure}
\usepackage{xcolor}
\usepackage[normalem]{ulem}
\usepackage[displaymath]{lineno}

\newcommand{\cL}{\mathcal{L}\,}

\newcommand{\grad}{\ensuremath{\bm{\nabla}}}

\newcommand{\bnabla}{\ensuremath{\bm{\nabla}}}

\newcommand{\xiv}{\boldsymbol{\xi}}

\newcommand{\Bv}{\mathbf{B}\,}
\newcommand{\cH}{\boldsymbol{\mathcal{H}}}

\newcommand{\ev}[1]{\,\hat{\boldsymbol{#1}}}

\newcommand{\cK}{\ensuremath{\mathcal{K}}}

\newcommand{\cP}{\boldsymbol{\mathcal{P}}}
\newcommand{\cT}{\boldsymbol{\mathcal{T}}}
\graphicspath{{./}{figures/}}

\received{August 15, 2022}
\revised{September 13, 2022}
\accepted{September 15, 2022}
\submitjournal{ApJ}

\shorttitle{Inferring magnetic fields using local mode-coupling}
\shortauthors{Srijan Bharati Das}

\begin{document}

\title{Recipe for inferring sub-surface solar magnetism via local mode-coupling using Slepian basis functions}

\correspondingauthor{Srijan Bharati Das}
\email{sbdas@princeton.edu}

\author[0000-0003-0896-7972]{Srijan Bharati Das}
\affiliation{Department of Geosciences \\
Princeton University \\
Princeton, New Jersey, USA}

\input{newcommands}


\begin{abstract}
Direct seismic imaging of sub-surface flow, sound-speed and magnetic field is crucial for predicting flux tube emergence on the solar surface, an important ingredient for space weather. The sensitivity of helioseismic mode-amplitude cross-correlation to $p$- and $f$-mode oscillations enable formal inversion of such sub-photospheric perturbations. It is well-known that such problems are written in the form of an integral equation that connects the perturbations to the observations via ``sensitivity kernels". While the sensitivity kernels for flow and sound-speed have been known for decades and have been used extensively, formulating kernels for general magnetic perturbations had been elusive. A recent study proposed sensitivity kernels for Lorentz-stresses corresponding to global magnetic fields of general geometry. The present study is devoted to proposing kernels for inferring Lorentz-stresses as well as the solenoidal magnetic field in a local patch on the Sun via Cartesian mode-coupling. Moreover, for the first time in solar physics, Slepian functions are employed to parameterize perturbations in the horizontal dimension. This is shown to increase the number of data constraints in the inverse problem, implying an increase in the precision of inferred parameters. This paves the path to reliably imaging sub-surface solar magnetic features in, e.g., supergranules, sunspots and (emerging) active regions.
\end{abstract}


\section{Introduction} \label{sec:intro}
The solar activity cycle has an approximate periodicity of eleven years. As the Sun goes from a solar minimum to a maximum, the number of active regions and sunspots goes from being nearly zero to around a couple hundred \citep[][and references therein]{hathaway_LRSP}. Formation and dissipation of these strongly magnetised patches is directly connected to the physical processes at play in the sub-surface layers, as detailed in the review by \cite{fan09}. Consequently, imaging sound-speed, flow and magnetic field in these layers is critical in understanding the solar cycle. While sound-speed and flow profiles have been widely investigated using traditional methods in local helioseismology \citep[see review by][and references therein]{gizon05}, direct seismic imaging of magnetic fields has not seen much success.

Attempts at studying sub-surface magnetic fields have mostly been in the area of sunspot seismology which dates back to \cite{Thomas82_sunspotseismology} reporting the splitting of the 5-minute oscillations. Using the Fourier-Hankel method, \cite{braun87} and \cite{bogdan93} observed absorption of $p$-mode power in sunspots. \cite{braun95} found strong evidence of ``mode-mixing" from correlation of phase shifts of the incident and scattered $p$-modes of adjacent radial orders, suggesting the need for using measurements that capture the scattering matrix. Conversion of $p$-modes to magneto-acoustic modes in a vertical flux tube was proposed in \cite{spruit91}. Similar efforts to explain the absorption of $p$-mode power continued in a series of notable studies that include but is not limited to \cite{spruit92}, \cite{cally93}, \cite{cally94}, \cite{cally97}, \cite{crouch05}, almost all of which use uniform and straight magnetic fields in modeling sunspots. \cite{cally05} adopted a perturbative approach while \cite{cally06} used a ray-theoretic approach to study conversion of acoustic modes to fast and slow magnetosonic modes when entering regions of strong uniform magnetic fields. Building on the ray-theoretic approach, \cite{Schunker06} studied the dependence of mode conversion on the ``attack angles" at which the acoustic waves impinge on magnetic fields. The reader is referred to \cite{Khomenko15} for a comprehensive review on similar studies. Another area of numerical effort includes the simulation of linearized wave-propagation through a magnetized background \citep[see][]{cameron08, cameron10, hanasoge_mag, schunker13}. 

Traditional methods in local helioseismology such as time-distance helioseismology \citep[][]{duvall}, helioseismic holography \citep[][]{lindsey97} and ring diagram analysis \citep[][]{hill88} have been implemented in inferring flow and sound-speed in and around sunspots and active regions. This is an extensive area of research with a controversial history and the reader is referred to the comprehensive study by \cite{gizon_etal_2009} and \cite{moradi09} and references therein. In particular, \cite{gizon_etal_2009} studied sub-surface flow and sound-speed perturbations around the NOAA active region 9787 using several methods in local helioseismology, and reported that the results could not be reconciled. Such disparity in results is attributed to the extreme sensitivity of each method to the complicated data analysis that precedes the inversion. Major breakthroughs in the area of numerically simulating a sunspot using realistic partial ionization and radiative transfer have been made in \cite{schussler}, \cite{khomenko06}, \cite{heinemann07}, \cite{rempel09}, \cite{felipe10}, \cite{rempel11_a}, \cite{rempel11_b}, \cite{rempel12}, \cite{rempel15}, \cite{Schmassmann22} among many others \citep[see][]{rempel_lrsp}.

The observed acoustic eigenstates of the Sun are largely consistent with the standard solar models such as Model~S \citep[][]{jcd}. These models are spherically symmetric, non-rotating, non-magnetic, isotropic and adiabatic. Observed departures of eigenfunctions and eigenfrequencies from the Model~S predictions can be explained due to the presence of unaccounted perturbations. The wavefields observed using Dopplergrams encode the coupling between normal-modes of the Sun due to the presence of perturbations such as rotation, flow fields, sound-speed anomaly, ellipticity and magnetic fields. Large-scale perturbations are inferred using global modes, arising from standing waves which traverse the bulk of the solar interior. Measurement of solar internal rotation marks one of the triumphs of global mode-coupling (GMC). \cite{howe09} summarizes the major breakthroughs in this area during the late 20$^{\mathrm{th}}$ and early 21$^{\mathrm{st}}$ century. In particular, the reader is referred to \cite{schou98} for details of multiple inversion methods for inferring solar internal rotation and their comparison. A weighted linear combination of internal perturbations $\delta p$, when integrated over the volume $V$, manifests itself as helioseismic measurements, say $\delta \omega$. An illustrative equation maybe written as
\begin{equation}
    \delta \omega = \int_V \mathcal{K}(x, y, z) \, \delta p(x, y, z) \, \rmd V \, .
\end{equation}
These weights $\mathcal{K}$, called ``sensitivity kernels", determine the sensitivity of observations to a perturbation of specific geometry and depth from the solar surface. In the absence of sensitivity kernels for global magnetic fields of a general configuration, earlier studies were limited to a simplified toroidal magnetic field \citep[][]{gough90, Antia2000, goode04, Kiefer17, Kiefer18}. This long-standing problem was addressed in \cite{Das2020} (hereafter D20), where sensitivity kernels for Lorentz-stress due to magnetic field of general geometry was presented. The resultant operator that describes the coupling of two modes closely spaced in frequency in the presence of magnetic field, called the coupling matrix, was demonstrated to be Hermitian. This is essential since an ideal MHD system is non-dissipative. The first part of this paper extends the GMC kernels in D20 to local mode-coupling (or LMC) kernels, for modes with angular degrees as high as 2000 \citep[as shown in][]{Mani22}, which are sensitive to sub-surface magnetic fields in localized patches on the Sun.

It can be shown that under the first-Born approximation, wavefield correlations are linearly related to perturbations via sensitivity kernels \citep[for local and global helioiseismic analysis the reader is referred to][]{woodard06, woodard14, woodard16}. Cartesian or local mode-coupling which was first formulated in \cite{woodard06} was successfully validated in \cite{Hanson21} by comparing supergranular power spectra with other local helioseismic results in \cite{gizon03} and \cite{langfellner18}. Very recently, LMC was set on firm footing by \cite{Mani22} where they inferred surface flows which were consistent with non-helioiseismic location correlation tracking \citep[][]{november88}. Access to the LMC magnetic field kernels would enable a collective inversion of magnetic field along with sound-speed and flow in quiet Sun patches and possibly also in strongly magnetized patches with sunspots and active regions. Explicit analytical expressions of kernels for both Lorentz-stress as well as a divergence-free magnetic field is presented in this study.

The second novelty of this study lies in the use of Slepian functions as an optimal basis in the horizontal plane in LMC. \cite{woodard06} and all other studies employing LMC have traditionally used a plane-wave basis in the horizontal coordinates to decompose the perturbations. Slepian functions, have until now, been used mostly in terrestrial applications such as investigating geodesy of polar caps \citep[][]{Slepian_polar_caps}, magnetization of Australian lithosphere \citep[][]{Slepian_australia} or melting of ice-sheets on Greenland \citep[][]{Slepian_Greenland}. They provide an effective basis in an arbitrary patch in spatial and spectral domain. The present study demonstrates the effectiveness of Slepian decomposition in solar applications by using HMI vector magnetograms of a sunspot NOAA 11084 and active region NOAA 12757 as well as a flow-map from local correlation tracking of an active region pre-emergence. It is shown that using a Slepian basis determined by the region of interest in physical and spectral space, increases the signal-to-noise ratio of the inferred model parameters by maximizing the number of data constraints used during inversion. The plane-wave kernels are shown to be trivially connected to Slepian kernels via a linear transformation, thereby allowing seamless adoption of these basis function in conjunction with the already existing machinery in LMC.

 The outline of this paper is as follows. The theory of Cartesian mode-coupling, its associated notations and results from D20 that are essential for this study are summarized in Section~\ref{sec: notation_and_formalism}. The individual components of kernels which are later used to build the final kernels for the six independent components of Lorentz-stress and a solenoidal magnetic field is listed in Section~\ref{sec: kernel_list}. The final kernels and linear inverse problem for Lorentz-stress in a plane-wave basis is described in Section~\ref{sec: BB_inversion}. Section~\ref{sec: slepian} introduces Slepian functions, adapts them to the linear inverse problem for Lorentz-stress components and demonstrates the decomposition of a sunspot and an active region in their respective Slepian basis. The kernels and non-linear inversion for a solenoidal magnetic field is described in Section~\ref{sec: inversion_B} and an ambiguity of joint inference of sound-speed anomaly and Lorentz-stress is illustrated in Section~\ref{sec: sound_speed_mag_anomaly}. Section~\ref{sec: discussion} summarizes the salient points of this study. For the sake of completeness, Appendix~\ref{sec: AppendixA} demonstrates the utility of Slepian decomposition for a non-magnetic perturbation --- a flow-map from averaged emerging active regions used in \cite{birch19}. Appendix~\ref{sec: slepian_functions_appendix} provides a brief outline of Slepian functions for the ease of reference and understanding of the reader.

\section{Notation and mathematical formalism} \label{sec: notation_and_formalism}

Calculations shown in this study are based on the original work by \cite{woodard06}, henceforth W06, where the formalism for wave propagation in a plane parallel atmosphere has been discussed. A brief description of the mathematical setup is presented in this section but the interested reader may refer to W06 for further details.

Wave propagation is considered in a ``propagating" Cartesian box defined by $x \in [-L_x,L_x], \, y \in [-L_y, L_y]$ and $z \in [z_{\mathrm{min}}, z_{\mathrm{max}}]$. Here, $(\ev{x}, \ev{y}, \ev{z})$ form a basis that tracks the local rotation with $\ev{z}$ pointing outwards from the solar surface, $\ev{y}$ pointing towards the solar North and $\ev{x} = \ev{y} \times \ev{z}$. The linearized wave equation in the frequency domain may be expressed as an eigenvalue problem
\begin{equation}
    \cL_0 \, \bfxi_{\alpha} = \rho_0 \, \omega_{\alpha}^2 \, \bfxi_{\alpha} \, ,
\end{equation}
where $\cL_0, \rho_0$ are the background zeroth-order wave operator and density, respectively. $(\omega_{\alpha}, \bfxi_{\alpha})$ are the unperturbed eigenfrequencies and eigenfunctions obtained by solving the eigenvalue problem using appropriate boundary conditions at $z_{\mathrm{min}}, z_{\mathrm{max}}$ as well as imposing periodicity at the horizontal boundaries \citep[see][]{birch}. The label $\alpha = (n, \bk)$ is the horizontal wave vector. Here, $n$ is the radial order indicating the number of nodes in $z$ and $\bk = (k_x, k_y)$. For the same $\bk$, the radial order is a monotonically increasing function of eigenfrequency $\omega$. Therefore, we may equivalently express $\bfxi_{\alpha}$ as  $\bfxi_{\bk}(\omega)$.

Since the linear operator $\cL_0$ describes wave propagation in a plane-parallel atmosphere, the eigenfunctions $\bfxi_{\bk}(\omega)$ may be decomposed in the basis of $e^{i \bk \cdot \bx}$
\begin{equation} \label{eq: xi_in_cart}
    \bfxi_{\bk}(\bx, z, \omega) = \left\{i\ev{k}\,H_k(z, \omega) + \ev{z} \, V_k (z, \omega) \right\} \, e^{i \, \bk \cdot \bx} \, ,
\end{equation}
where $H_k(z, \omega)$ and $V_k(z, \omega)$ are the real-valued horizontal and vertical eigenfunctions, respectively. For convenience of understanding, analogies of LMC with GMC are listed. $\ell, m$ in GMC encode the number of nodes in latitude and longitude. Similarly, the wave vector $\bk = (k_x, k_y)$ count the number of nodes in $x$ and $y$. The radial direction in GMC $r$ is analogous to depth $z$ in LMC. The global eigenfunctions in GMC is expressed in the basis of vector spherical harmonics $Y_{\ell, m} \ev{r}$ (which become $\ev{z} \, e^{i \bk \cdot \bx}$ in LMC) and $\bnabla_{H} Y_{\ell,m}$ (which become $i \bk \, e^{i \bk \cdot \bx}$ in LMC). The global eigenfunctions $U_{n \ell}(r)$ and $V_{n\ell}(r)$ which were functions of radius $r$ are analogous to the local cartesian eigenfunctions $V_{k}(z, \omega)$ and $H_{k}(z, \omega)$ which are functions of depth $z$. While the above helps in building an intuitive parallel between the local and global cases, the reader is encouraged to see Appendix~A of \cite{birch} for details of variable separation leading to the form of eigenfunctions as in Eqn.~(\ref{eq: xi_in_cart}).

The complex mode amplitude $\varphi_{\bk}^{\omega}$ is an observable measured from dopplergrams. The total wavefield in background model $\bfxi(\omega)$ can be expressed as a linear combination of the unperturbed eigenfunctions 
\begin{equation}
    \bfxi(\bx, z, \omega) = \sum_{\bk} \varphi_{\bk}^{\omega} \, \bfxi_{\bk}(\bx, z, \omega) \, .
\end{equation}  In the presence of linear perturbations in the background model, the linear wave operator changes by $\delta \cL$ which causes the wavefield to change to $\bfxi \xrightarrow{} \bfxi + \delta \bfxi$ where
\begin{equation}
    \delta \bfxi = \sum_{\bk} \delta \varphi_{\bk}^{\omega} \, \bfxi_{\bk} \, .
\end{equation}
As initially laid out in W06 and later used in \cite{Hanson21} and \cite{Mani22}, in the LMC context, it can be shown that the expectation value of these mode amplitude cross-spectra $\langle\varphi_{\bk'}^{\omega' *} \, \delta\varphi_{\bk}^{\omega} + \delta \varphi_{\bk'}^{\omega' *} \, \varphi_{\bk}^{\omega}\rangle$ can been used as an observable since it is linearly related to the coupling matrix $\Lambda_{k'k} = \int_{V} \rho \, \bfxi_{\bk'}^* \cdot \delta \cL \bfxi_{\bk} \rmd V$ as
\begin{equation} \label{eqn: mode_coupling}
    \langle \varphi_{k'}^{\omega'} \delta\varphi_k^{\omega *} + \delta \varphi_{k'}^{\omega'} \varphi_k^{\omega *} \rangle = (N_{k'} R_k^{\omega *} |R_{k'}^{\omega'}|^2 + N_{k} R_{k'}^{\omega'} |R_{k}^{\omega}|^2) \Lambda_{k'k} \, ,
\end{equation}
where $\langle A \rangle$ represents the statistical expectation value of parameter $A$, $N_k$ is the mode-amplitude normalization and $R_k^{\omega} = (\omega_k^2 - \omega^2 - i \omega \, \gamma_k/2)^{-1}$, where $\gamma_k$ is the mode linewidth. When considering magnetic perturbations due to the presence of magnetic field $\bfB = \bfB_0 + \bfB_1$ in an ideal MHD limit, it can be shown that \citep[refer to Chapter 6 in ][]{goedbloed2004}
\begin{equation}
    \delta \cL = \mathbf{B}_0 \times (\bnabla \times \mathbf{B}_1) - (\bnabla \times \mathbf{B}_0) \times \mathbf{B}_1 - \bnabla [\bfxi \cdot (\mathbf{j}_0 \times \mathbf{B}_0)] \, ,
\end{equation}
where $\bfB_0$ is the zeroth order field, $\mathbf{j}_0 = \bnabla \times \bfB_0$ is the current density and $\mathbf{B}_1 = \bnabla \times (\xiv \times \mathbf{B}_0)$ is the first order correction. We drop the subscript `0' in the rest of the paper and $\bfB$ shall be assumed to imply the zeroth-order magnetic field. The second-rank Lorentz-stress $\bfB \bfB$ is the natural choice of physical quantity for inversions. This is because for any two vectors $\bfA$ and $\bfC$, the following vector calculus identity holds true $\bnabla(\bfA \cdot \bfC) = \bfA \times (\bnabla \times \bfC) + \bfC \times (\bnabla \times \bfA) + (\bfA \cdot \bnabla) \, \bfC + (\bfC \cdot \bnabla) \, \bfA$. Using this it can be shown that
\begin{equation}
    \bfj \times \bfB = \bnabla \cdot \left(\bfB \, \bfB -  \frac{B^2}{2}\, \bfI \right) \, .
\end{equation}
As shown in Appendix C of D20, carrying out sequential integration by parts, $\bfB \bfB$ can be disentangled from the kernel as the invertible parameter. The resultant coupling matrix can then be expressed as
\begin{equation} \label{eq: coupling_matrix}
    \Lambda_{\bk'\bk}^{\mathrm{mag}} = \int_{V} \bfB \bfB : \boldsymbol{\cK}_{\bk'\bk} \, \mathrm{d}V \, ,
\end{equation}where
\begin{eqnarray} \label{eqn: BB_kernel}
    \boldsymbol{\cK}_{\bk'\bk} &=& \frac{1}{4\pi} \left\{\tfrac{1}{2}\left[\bnabla \xiv_k \cdot (\grad \xiv_{k'}^*)^T + \bnabla \xiv_{k'}^* \cdot (\bnabla \xiv_k)^T \right] + \tfrac{1}{2}\left( \bnabla \xiv_{k'}^* \cdot \bnabla \xiv_k + \bnabla \xiv_k \cdot \bnabla \xiv_{k'}^*\right) + \boldsymbol{I}\, \bnabla \cdot \xiv_{k'}^* \bnabla \cdot \xiv_k \right. \nonumber\\
    &+& \left. \tfrac{1}{2}\left( \xiv_{k'}^*\cdot \bnabla \bnabla \xiv_k + \xiv_k \cdot \bnabla \bnabla \xiv_{k'}^* \right) 
    - \tfrac{1}{2}\left( \xiv_k \bnabla \bnabla \cdot \xiv_{k'}^* + \xiv_{k'}^* \bnabla \bnabla \cdot \xiv_k \right) - \tfrac{3}{2} \left( \bnabla \xiv_k \bnabla \cdot \xiv_{k'}^* + \bnabla \xiv_{k'}^* \bnabla \cdot \xiv_k \right)  \right\} \, ,
\end{eqnarray}
along with the following boundary terms
\begin{equation} \label{eq: mag_surface_terms}
   \frac{1}{4\pi} \int_{\Sigma} d \Sigma \,\hat{\mathbf{n}} \cdot \big[ \xiv^*_{k'}\cdot(\bnabla \, \Bv) \, \Bv \cdot \xiv_{k} + \Bv \Bv \cdot \xiv_{k} (\bnabla \cdot \xiv^*_{k'}) - \tfrac{1}{2}\Bv \Bv : \bnabla(\xiv_{k} \, \xiv^*_{k'})\big]+ (\xiv_k \leftrightarrow \xiv^*_{k'}) \,.
\end{equation}
In Eqn.~(\ref{eq: coupling_matrix}), the ``:" operator is used to indicate matrix contraction such that, for two rectangular matrices $\bfA$ and $\bfC$ with compatible dimensions, the operation $\bfA : \bfC = A_{ij} \, C_{ji}$ (note that summation is implied on repeated indices here). The double-sided arrow in Eqn.~(\ref{eq: mag_surface_terms}) is used as a short-hand to imply the swapping of $k$ and $k'$ in the expressions for eigenfunctions followed by taking a complex conjugate. The above expressions are for a generic coordinate system. D20 calculated the analytical expressions in a spherical polar coordinate system. This study is dedicated to finding the explicit analytical expressions for the Lorentz-stress kernel in a Cartesian geometry.

\section{Lorentz-stress sensitivity kernels in Cartesian geometry} \label{sec: kernel_list}

Using Eqn.~(\ref{eq: xi_in_cart}) and $\bnabla = \ev{z} \, \partial_z + \bnabla_{H}$ it is conceptually straightforward to find the explicit form of the kernel $\boldsymbol{\cK}_{\bk\bk'}$ in terms of the horizontal and vertical eigenfunctions $H_k(\omega)$ and $V_{k}(\omega)$ respectively. Substituting these in Eqn.~(\ref{eqn: BB_kernel}), it can be shown that the Lorentz-stress kernels $\boldsymbol{\cK}_{\bk\bk'}$ can be expanded into the following tensor components 
\begin{equation} \label{eq: kernel_comps}
\begin{split}
    e^{i (\bk' - \bk) \cdot \bx} \, \boldsymbol{\cK}_{\bk\bk'} =& \, \cK_{\bk\bk'}^{xx} \ev{x} \ev{x} + \cK_{\bk\bk'}^{yy} \ev{y} \ev{y} + \cK_{\bk\bk'}^{zz} \ev{z} \ev{z} + \cK_{\bk\bk'}^{zk} \ev{z} \ev{k} + \cK_{\bk\bk'}^{zk'} \ev{z} \ev{k}' + \cK_{\bk\bk'}^{kz} \ev{k} \ev{z} \\
    &+ \cK_{\bk\bk'}^{k'z} \ev{k}' \ev{z} + \cK_{\bk\bk'}^{kk} \ev{k} \ev{k} + \cK_{\bk\bk'}^{k'k'} \ev{k}' \ev{k}' + \cK_{\bk\bk'}^{kk'} \ev{k} \ev{k}' + \cK_{\bk\bk'}^{k'k} \ev{k}' \ev{k} \, ,
\end{split}
\end{equation}
where
\begin{align}
\begin{split}
    \cK_{\bk\bk'}^{xx} = & \, \dot{V}_k \, \dot{V}_{k'} + k\,k'\,H_{k}\,H_{k'} - k'\, \dot{V}_{k} \, H_{k'} - k \, \dot{V}_{k'} \, H_{k} \, ,
\end{split}\\
\begin{split}
    \cK_{\bk\bk'}^{yy} = & \, \dot{V}_k \, \dot{V}_{k'} + k\,k'\,H_{k}\,H_{k'} - k'\, \dot{V}_{k} \, H_{k'} - k \, \dot{V}_{k'} \, H_{k} \, ,
\end{split}\\
\begin{split}
    \cK_{\bk\bk'}^{zz} = & \, \left\{\dot{H}_k \, \dot{H}_{k'} + \frac{1}{2} \left(k \, \dot{V}_k \, H_{k'} + k' \, \dot{V}_{k'} \, H_{k} + k \, V_k \, \dot{H}_{k'} + k' \, V_{k'} \, \dot{H}_k\right) \right\} \, (\ev{k} \cdot \ev{k}') \\ 
    &+ \left\{ k\,k'\,H_{k}\,H_{k'} + \frac{1}{2} \left(k'\, \dot{V}_{k} \, H_{k'} + k \, \dot{V}_{k'} \, H_{k} + k' \, V_k \, \dot{H}_{k'} + k \, V_{k'} \, \dot{H}_k\right)\right\} \, ,
\end{split} \\
\begin{split} \label{eq: Kzk}
    \cK_{\bk\bk'}^{zk} =& \, i \left(k \, H_{k} \, \dot{H}_{k'} + \frac{k}{2} \, \dot{H}_k \, H_{k'}\right) (\ev{k} \cdot \ev{k}') \\
    &+ \frac{i}{2} \left(k \, V_k \, \dot{V}_{k'}  + \ddot{H}_k  \, V_{k'} -  \dot{H}_k \, \dot{V}_{k'} +  k' \, \dot{H}_k \, H_{k'} - k \, V_{k'} \, \dot{V}_k + k^2 \, V_{k'} \, H_k \right) \, ,
\end{split} \\
\begin{split}
    \cK_{\bk\bk'}^{zk'} =& \, -i \left(k' \, H_{k'} \, \dot{H}_{k} + \frac{k'}{2} \, \dot{H}_{k'} \, H_{k}\right) (\ev{k} \cdot \ev{k}') \\
    &- \frac{i}{2} \left(k' \, V_{k'} \, \dot{V}_{k}  + \ddot{H}_{k'}  \, V_{k} -  \dot{H}_{k'} \, \dot{V}_{k} +  k \, \dot{H}_{k'} \, H_{k} - k' \, V_{k} \, \dot{V}_{k'} + k'^2 \, V_{k} \, H_{k'} \right) \, ,
\end{split} \\
\begin{split}
    \cK_{\bk\bk'}^{kz} =& \, \frac{i}{2} \left( k \, H_k \, \dot{H}_{k'} +  k^2 \, V_k \, H_{k'} + k \, k' \, V_{k'} \, H_k\right) (\ev{k} \cdot \ev{k}') \\ &+ \frac{i}{2} \left( k \, \dot{V}_k \, V_{k'} - k \, \dot{V}_{k'} \, V_k + 3 k \, k' \, V_k \, H_{k'} + k' \, H_k \, \dot{H}_{k'} - H_k \, \ddot{V}_{k'} \right) \, ,
\end{split} \\
\begin{split} \label{eq: Kk_z}
    \cK_{\bk\bk'}^{k'z} =& \, -\frac{i}{2} \left( k' \, H_{k'} \, \dot{H}_{k} +  k'^2 \, V_{k'} \, H_{k} + k' \, k \, V_{k} \, H_{k'}\right) (\ev{k} \cdot \ev{k}') \\  &- \frac{i}{2} \left( k' \, \dot{V}_{k'} \, V_{k} - k' \, \dot{V}_{k} \, V_{k'} + 3 k' \, k \, V_{k'} \, H_{k} + k \, H_{k'} \, \dot{H}_{k} - H_{k'} \, \ddot{V}_{k} \right) \, ,
\end{split} \\
\begin{split}
    \cK_{\bk\bk'}^{kk} =& \, \frac{3}{2} \left( k \, H_k \, \dot{V}_{k'} - k \, k' \, H_k \, H_{k'} \right) - \frac{1}{2} \left\{ k^2 \, H_k \, H_{k'} (\ev{k} \cdot \ev{k}') + k \, \dot{H}_k \, V_{k'} \right\} \, ,
\end{split} \\
\begin{split}
    \cK_{\bk\bk'}^{k'k'} =& \, \frac{3}{2} \left( k' \, H_{k'} \, \dot{V}_k - k \, k' \, H_{k} \, H_{k'} \right) - \frac{1}{2} \left\{ k'^2 \, H_k \, H_{k'} (\ev{k} \cdot \ev{k}') + k' \, V_k \, \dot{H}_{k'} \right\} \, ,
\end{split} \\
\begin{split}
    \cK_{\bk\bk'}^{kk'} =& \, k \, k' \, H_k \, H_{k'} \, (\ev{k} \cdot \ev{k}') + \frac{1}{2} \left\{ k \, k' \, V_k \, V_{k'} -  k' \, H_k \dot{V}_{k'} + k'^2 \, H_k \, H_{k'} + k \, \dot{H}_{k'} \, V_k \right\} \, ,
\end{split} \\
\begin{split}
    \cK_{\bk\bk'}^{k'k} =& \, k \, k' \, H_k \, H_{k'} \, (\ev{k} \cdot \ev{k}') + \frac{1}{2} \left\{ k \, k' \, V_k \, V_{k'} -  k \, H_{k'} \dot{V}_{k} + k^2 \, H_{k'} \, H_{k} + k' \, \dot{H}_{k} \, V_{k'} \right\} \, .
\end{split}
\end{align}

\begin{figure}
    \centering
    \includegraphics[width=\textwidth]{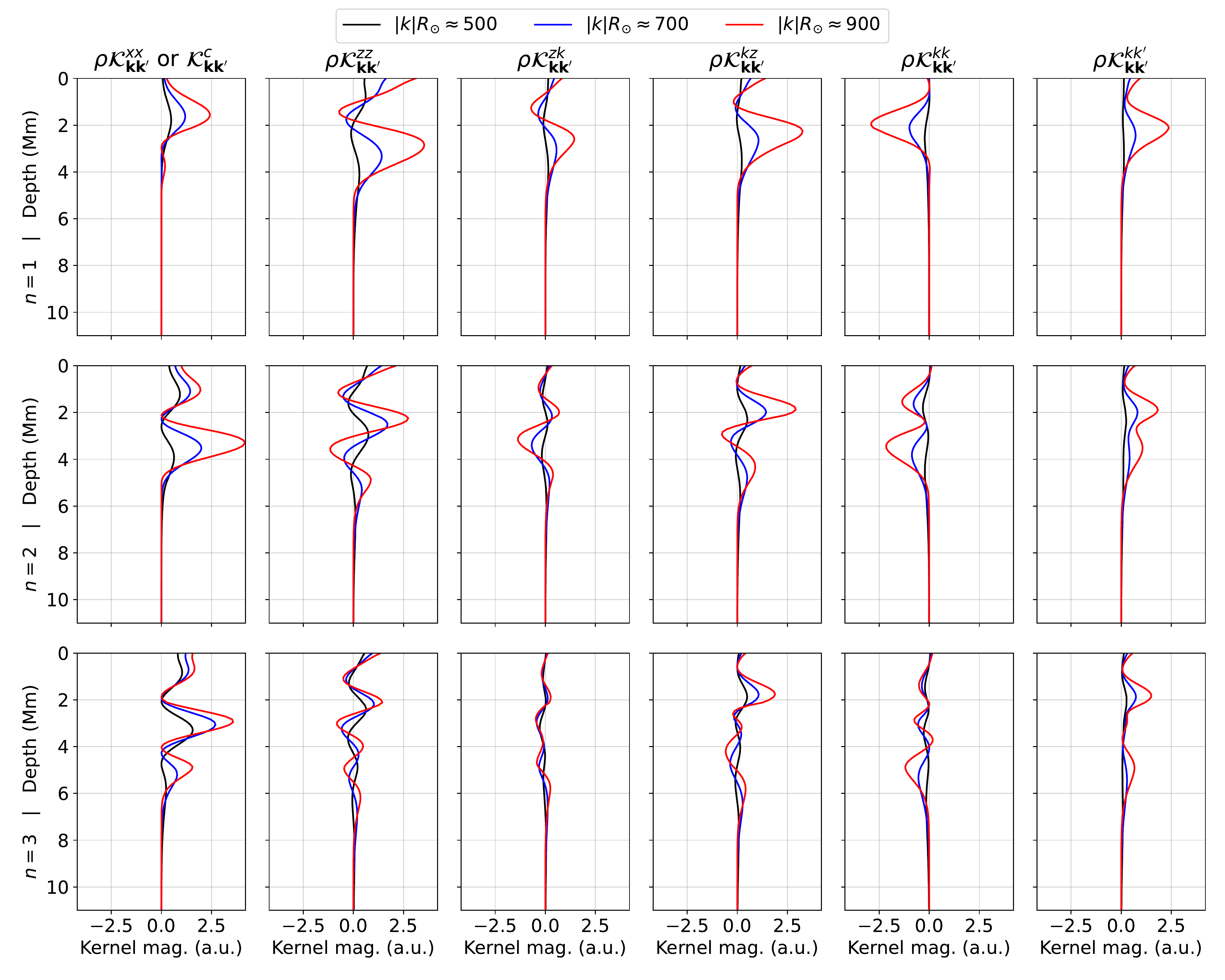}
    \caption{Components of the distinct Lorentz-stress kernels present in Eqn.~(\ref{eq: kernel_comps}), scaled by density $\rho(z)$, plotted from the surface down to a depth of 11Mm. The $x$-axis is the kernel magnitude in arbitrary units (a.u.) and the $y$-axis is the depth $z$ starting at the surface at $z=0$ and going deeper for larger values of $z$. As shown in Sec.~\ref{sec: sound_speed_mag_anomaly}, the \textit{top-left} panel is also the sound-speed kernel $\cK_{\bk\bk'}^c$ plotted in the same arbitrary units for comparison with the Lorentz-stress kernels. Only the self-coupling kernels are plotted where $\bk = \bk'$. The \textit{black, blue} and $\textit{red}$ lines are in increasing order of wave-number magnitude. $\bk \, R_{\odot}$ has the value (a) (-471.2, -168.3) for the \textit{black} line, (b) (-650.7, -258.1) for the \textit{blue} line, and (c) (-673.2, -594.6) for the \textit{red} line. The rows are arranged in ascending radial order $n = 1, 2, 3$ of the eigenfunctions used.}
    \label{fig:lorentz_kernels}
\end{figure}
Here, the partial derivatives in depth, $\partial_z$, are denoted by overdots, e.g., $\partial_z H_k \equiv \dot{H}_k$. As mentioned before, since $H_k$ and $V_k$ are real-valued functions, it may be noted that
\begin{itemize}
    \item $\cK_{\bk\bk'}^{xx} = \cK_{\bk\bk'}^{yy}$,
    \item $\cK_{\bk\bk'}^{xx}, \cK_{\bk\bk'}^{yy}$ and $K_{\bk\bk'}^{zz}$ are symmetric in $k$ and $k'$,
    \item only $\cK_{\bk\bk'}^{zk}, \cK_{\bk\bk'}^{zk'}, \cK_{\bk\bk'}^{kz}$ and $\cK_{\bk\bk'}^{k'z}$ are purely imaginary while the rest of the component are purely real,
    \item for the imaginary components: $\cK_{\bk\bk'}^{zk} = \cK_{\bk\bk'}^{zk'*} (k \leftrightarrow k')$ and $\cK_{\bk\bk'}^{kz} = \cK_{\bk\bk'}^{k'z*} (k \leftrightarrow k')$,
    \item for the real components: $\cK_{\bk\bk'}^{kk} = \cK_{\bk\bk'}^{k'k'}(k \leftrightarrow k')$ and $\cK_{\bk\bk'}^{kk'} = \cK_{\bk\bk'}^{k'k}(k \leftrightarrow k')$.
\end{itemize}
The double-sided arrows in the tabulated relations are used as a short-hand to imply the swapping of $k$ and $k'$ in the expressions for kernels. These symmetry relations may be used to demonstrate that the coupling matrix $\Lambda_{k'k}$ is Hermitian and hence, the eigenfrequencies are real. Therefore, there is no damping or instability of modes due to these magnetic perturbations in the ideal MHD limit.

Figure~\ref{fig:lorentz_kernels} shows magnitudes of the 6 distinct components of the kernels listed above as a function of depth $z$ up~to 11~Mm from $z = 0$. Three different radial orders $n= 1, 2, 3$ are plotted in different rows and the kernel components are plotted in subsequent columns. Uniform extent in $x \in [-2.5, 2.5]$ (in arbitrary units, or ``a.u.") is used for all the plots for ease of comparison. The plotted kernels are for self-coupling, i.e., $\bk = \bk'$. Using the symmetry relations listed above, only the six independent components $\cK_{\bk\bk'}^{xx}$, $\cK_{\bk\bk'}^{zz}$, $\cK_{\bk\bk'}^{zk}$, $\cK_{\bk\bk'}^{kz}$, $\cK_{\bk\bk'}^{kk}$ and $\cK_{\bk\bk'}^{kk'}$ are presented in successive columns. As shown explicitly in Eqns.~(\ref{eq: Kzk})-(\ref{eq: Kk_z}), the components $\cK_{\bk\bk'}^{zk}$ and $\cK_{\bk\bk'}^{kz}$ are purely imaginary. Only the magnitude of the imaginary part for these kernels are plotted. The \textit{black}, \textit{blue} and \textit{red} lines correspond to the kernels built from modes with the absolute value of horizontal wavenumber $|k| R_{\odot} \approx 500, 700, 900$ respectively. More specifically, $(k_x R_{\odot}, k_y R_{\odot}) = (-471.2, -168.3)$ for the kernels in plotted \textit{black}. For the kernels in \textit{blue}, $(k_x R_{\odot}, k_y R_{\odot}) = (-650.7, -258.1)$ and for those in \textit{red}, $(k_x R_{\odot}, k_y R_{\odot}) = (-673.2, -594.6)$. As visible across all components, kernels corresponding to lower radial orders $n$ are sensitive to a shallower depth as compared to kernels built from larger radial orders. Closer inspection shows that kernels build from modes with higher $|k| R_{\odot}$, have stronger sensitivities but die out quicker than kernels built from lower $|k| R_{\odot}$ modes. All the kernel components have roughly the same strength, except perhaps for $\mathcal{K}_{\bk\bk'}^{zk}$ which appears to be slightly smaller in magnitude. For the purely imaginary kernels $\mathcal{K}_{\bk\bk'}^{zk}$ and $\mathcal{K}_{\bk\bk'}^{kz}$, only their respective magnitudes are plotted.

\section{Formulating the inverse problem} \label{sec: inversion}
Inferring solar internal magnetic fields via an inverse problem involves formulating the forward problem that encodes the physics of wave-propagation in a magnetized medium. This includes modeling the magnetic perturbations as unknowns in an optimal basis and finding sensitivity kernels associated to these unknowns. The following subsections are arranged such that the reader is first introduced to the kernels for the six independent components of the Lorentz-stress in Section~(\ref{sec: BB_inversion}). This is according to the standard prescription of LMC in a plane-wave basis. Subsequently, the Slepian functions are introduced in Section~(\ref{sec: slepian}) and are used to decompose and reconstruct surface Lorentz-stresses for the sunspot NOAA 11084 and the active region NOAA 12757. This section ends with constructing the Slepian kernels (new kernels in the Slepian basis) using the plane-wave kernels found in Section~(\ref{sec: BB_inversion}). Finally, Section~(\ref{sec: inversion_B}) lays out the forward problem assuming the divergence-free magnetic field $\bfB$ instead of the Lorentz-stress $\bfB \bfB$ as the model parameter. This is shown to lead to a non-linear inverse problem which is not suitable for using the Slepian basis.

\subsection{Linear inversion: With Lorentz-stress $\bfB \bfB$ as the model parameter} \label{sec: BB_inversion}
In the Cartesian geometry, the second rank Lorentz-stress tensor $\cH = \bfB \bfB$ can be written as
\begin{equation} \label{eq: H_comps}
    \cH = \mathcal{H}^{xx} \, \ev{x}\ev{x} + \mathcal{H}^{yy} \, \ev{y}\ev{y} + \mathcal{H}^{zz}\, \ev{z}\ev{z} + \mathcal{H}^{xy} \, (\ev{x}\ev{y} + \ev{y}\ev{x}) + \mathcal{H}^{xz} \, (\ev{x}\ev{z} + \ev{z}\ev{x}) + \mathcal{H}^{yz} \, (\ev{y}\ev{z} + \ev{z}\ev{y}) \, ,
\end{equation}
where the six independent components for inversion are $\mathcal{H}^{xx} = B_x^2, \, \mathcal{H}^{yy} = B_y^2, \, \mathcal{H}^{zz} = B_z^2, \, \mathcal{H}^{xy} = B_x \, B_y, \, \mathcal{H}^{xz} = B_x \, B_z$ and $\mathcal{H}^{yz} = B_y \, B_z$. In order to find the coupling matrix $\Lambda_{\bk\bk'}$, it is necessary to carry out a contraction between the expression of $\bfB \bfB$ given in Eqn.~(\ref{eq: H_comps}) with that of $\boldsymbol{\cK}_{\bk\bk'}$ in Eqn.~(\ref{eq: kernel_comps}). Defining $\ev{k} \cdot \ev{x} = \mu_{\bk \bx}$ and $\ev{k} \cdot \ev{y} = \mu_{\bk \by}$ we get
\begin{equation}
    \bfB \bfB : \boldsymbol{\cK}_{\bk\bk'} = \left(\mathcal{B}_{xx} \, \mathcal{H}^{xx} + \mathcal{B}_{yy} \, \mathcal{H}^{yy} + \mathcal{B}_{zz} \, \mathcal{H}^{zz} + \mathcal{B}_{xy} \, \mathcal{H}^{xy} + \mathcal{B}_{xz} \, \mathcal{H}^{xz} + \mathcal{B}_{yz} \, \mathcal{H}^{yz} \right) \, e^{i (\bk - \bk') \cdot \bx}\, ,
\end{equation}
where the six individual components of the kernel tensor $\boldsymbol{\mathcal{B}}$, which corresponds to sensitivity kernels for the six independent components of the Lorentz-stress tensor $\cH$, are listed below:
\begin{eqnarray}
    \mathcal{B}_{xx}(\bk,\bk', z) &=& \quad \cK_{\bk\bk'}^{xx} + \mu_{\bk \bx} \, \mu_{\bk \bx} \, \cK_{\bk\bk'}^{kk} + \mu_{\bk' \bx} \, \mu_{\bk' \bx} \, \cK_{\bk\bk'}^{k'k'} + \mu_{\bk \bx} \, \mu_{\bk' \bx} \, \cK_{\bk\bk'}^{kk'} + \mu_{\bk' \bx} \, \mu_{\bk \bx} \, \cK_{\bk\bk'}^{k'k} \, , \label{eq: effective_BB_kernels_1}\\
    \mathcal{B}_{yy}(\bk,\bk', z) &=& \quad \cK_{\bk\bk'}^{yy} + \mu_{\bk \by} \, \mu_{\bk \by} \, \cK_{\bk\bk'}^{kk} + \mu_{\bk' \by} \, \mu_{\bk' \by} \, \cK_{\bk\bk'}^{k'k'} + \mu_{\bk \by} \, \mu_{\bk' \by} \, \cK_{\bk\bk'}^{kk'} + \mu_{\bk' \by} \, \mu_{\bk \by} \, \cK_{\bk\bk'}^{k'k} \, , \\
    \mathcal{B}_{zz}(\bk,\bk', z) &=& \quad \cK_{\bk\bk'}^{zz} \, , \\ 
    \mathcal{B}_{xy}(\bk,\bk', z) &=& \quad (\mu_{\bk \by} \, \mu_{\bk \bx} + \mu_{\bk \bx} \, \mu_{\bk \by}) \, \cK_{\bk\bk'}^{kk} + (\mu_{\bk' \by} \, \mu_{\bk' \bx} + \mu_{\bk' \bx} \, \mu_{\bk' \by}) \, \cK_{\bk\bk'}^{k'k'} \nonumber\\
    && + \, (\mu_{\bk' \by} \, \mu_{\bk \bx} + \mu_{\bk' \bx} \, \mu_{\bk \by}) \, \cK_{\bk\bk'}^{kk'} + (\mu_{\bk \by} \, \mu_{\bk' \bx} + \mu_{\bk \bx} \, \mu_{\bk' \by}) \, \cK_{\bk\bk'}^{k'k}  \, , \\
    \mathcal{B}_{xz}(\bk,\bk', z) &=& \quad \mu_{\bk \bx} \, (\cK_{\bk\bk'}^{zk} + \cK_{\bk\bk'}^{kz}) + \mu_{\bk' \bx} \, (\cK_{\bk\bk'}^{zk'} + \cK_{\bk\bk'}^{k'z}) \, , \\
    \mathcal{B}_{yz}(\bk,\bk', z) &=& \quad \mu_{\bk \by} \, (\cK_{\bk\bk'}^{zk} + \cK_{\bk\bk'}^{kz}) + \mu_{\bk' \by} \, (\cK_{\bk\bk'}^{zk'} + \cK_{\bk\bk'}^{k'z}) \, . \label{eq: effective_BB_kernels_6}
\end{eqnarray}
Fig.~(\ref{fig:lorentz_kernels}) shows that the strength of almost all the kernel components are similar. Investigating the terms containing projections $\mu_{\bk \bx}, \mu_{\bk' \bx}, \mu_{\bk \by}$ and $\mu_{\bk' \by}$ in Eqns.~(\ref{eq: effective_BB_kernels_1})-(\ref{eq: effective_BB_kernels_6}) shows that $\mathcal{B}_{xx} \, (\sim \cK^{xx}_{\bk\bk'}), \, \mathcal{B}_{yy} \, (\sim \cK^{yy}_{\bk\bk'})$ and $\mathcal{B}_{zz}$ have the strongest sensitivity. This implies that the isotropic Lorentz-stress components $B^2_x, \, B^2_y$ and $B^2_z$, responsible for magnetic pressure, have the strongest seismic sensitivity. $B_x \, B_y$ has the lowest sensitivity due to the presence of cross projections such as $\mu_{\bk \bx} \, \mu_{\bk \by}$ (and other combinations of $k, k'$) in $\mathcal{B}_{xy}$.

To setup the inversion, the horizontal profile of $\cH$ may be decomposed into the basis of $e^{i \bq \cdot \bx}$, where $\bq$ is a label for the horizontal wavenumber of the perturbation. The depth variation of each component of the $\cH$ tensor may be parameterized in a B-splines basis formed by piecewise-polynomials which are weighted by coefficients, say $c_j$. Here $j$ is the knot location in depth where local support is provided by the spline function $f^j(z)$. This parameterization may be illustrated by decomposing $\cH$ in the following two steps
\begin{eqnarray}
    \cH(x, y, z) &=& \int \cH_{\bq}(z) \, e^{i \bq \cdot \bx} \rmd \bq \, , \\
    \cH(x, y, z) &=& \sum_j \int f^j(z) \, \cH_{j \bq} \, e^{i \bq \cdot \bx} \rmd \bq \, .
\end{eqnarray}
Here, $\cH_{j \bq} = \left(\mathcal{H}_{j \bq}^{xx}, \,  \mathcal{H}_{j \bq}^{yy}, \, \mathcal{H}_{j \bq}^{zz}, \, \mathcal{H}_{j \bq}^{xy}, \, \mathcal{H}_{j \bq}^{xz}, \, \mathcal{H}_{j \bq}^{yz}\right)$. Finally, the statement for inverse problem becomes
\begin{equation} \label{eq: coupling_matrix_BB}
    \Lambda_{\bk\bk'}^{\mathrm{mag}} = \sum_{\gamma,\delta} \sum_j \mathcal{B}_{\gamma \delta}^j(\bk,\bk') \, \mathcal{H}^{\gamma \delta}_{j \bq} \, ,
\end{equation}
where $\bk' =\bk + \bq$, the labels $\gamma, \delta = \{x, y, z\}$ and $\mathcal{B}_{\gamma \delta}^j(\bk,\bk') = \int f^j(z) \, \mathcal{B}_{\gamma \delta}(\bk,\bk', z) \, \rmd z$. Consequently, the above equation implies that for a particular model parameter $\mathcal{H}_{j_0 \bq_0}^{\gamma_0 \delta_0}$ of Lorentz-stress given by the tensor component $(\gamma_0, \delta_0)$ at a depth location given by knot $j_0$, the wavenumber $\bq_0$ of perturbation depends on the pairs of interacting modes, say $\bk_0$ and $\bk'_0$. Following the above selection rule, only the small subset of modes $\bk_0 \in \mathbf{K}$ and $\bk'_0 \in \mathbf{K}'$ that satisfy $\bq_0 = \bk'_0 - \bk_0$ will constrain the model parameter $\cH_{j_0\bq_0}^{\gamma_0 \delta_0}$. Here $\mathbf{K}$ and $\mathbf{K}'$ are subsets of the total mode superset $\mathbf{K}_{\mathrm{super}}$ containing all the observed modes in the coupling matrix. In contrast to this, Section~\ref{sec: slepian} discusses how, when expressed in the Slepian basis, the model parameters are constrained collectively by all the available modes in $\mathbf{K}_{\mathrm{super}}$ as opposed to a subset of it. The inverse problem statement in Eqn.~(\ref{eq: coupling_matrix_BB}), connects the model parameters $\mathcal{H}^{\gamma \delta}_{j \bq}$ to the observables $\Lambda_{\bk\bk'}^{\mathrm{mag}}$ via weights known from this study given by the kernel components $\mathcal{B}_{\gamma \delta}^j(\bk,\bk')$. So, $\Lambda_{\bk\bk'}^{\mathrm{mag}}$ is linearly related to $\mathcal{H}^{\gamma \delta}_{j \bq}$, and this forms a simple linear inverse problem solvable using methods like Regularized Least-Square (RLS) or Subtractive Optimally Localized Averages \citep[SOLA, see][]{pijpers}.

\subsection{Decomposing perturbations in the horizontal plane using Slepian basis} \label{sec: slepian}

Efficient parameterization of a perturbation is crucial for minimizing the number of unknowns being inferred and thereby increasing the effective number of constraints per model parameter. Cartesian mode-coupling has traditionally used a plane-wave basis for parameterizing perturbations in the horizontal plane. While all previous sections in this paper have adopted the same template, it can be shown that using Slepian functions as basis functions increases the effective number of data constraints for the model parameters. Using a Slepian basis also comes with the advantage of focusing only on the region of interest thereby efficiently eliminating noise from adjacent pixels in the data-cube (this is most clearly seen in the example in Appendix~\ref{sec: AppendixA}). This section demonstrates the use of Slepian functions in helioseismology when dealing with magnetic perturbations in sunspots and active regions. Appendix~\ref{sec: AppendixA} demonstrates the same for flow perturbation in emerging active region and Appendix~\ref{sec: slepian_functions_appendix} outlines the basic ideas useful in understanding Slepian functions in a Cartesian plane. Using Slepian functions render similar advantages when inferring flow, sound-speed and magnetization in supergranules, which will be shown in a future paper.

\begin{figure}
\centering
  \includegraphics[width=\textwidth]{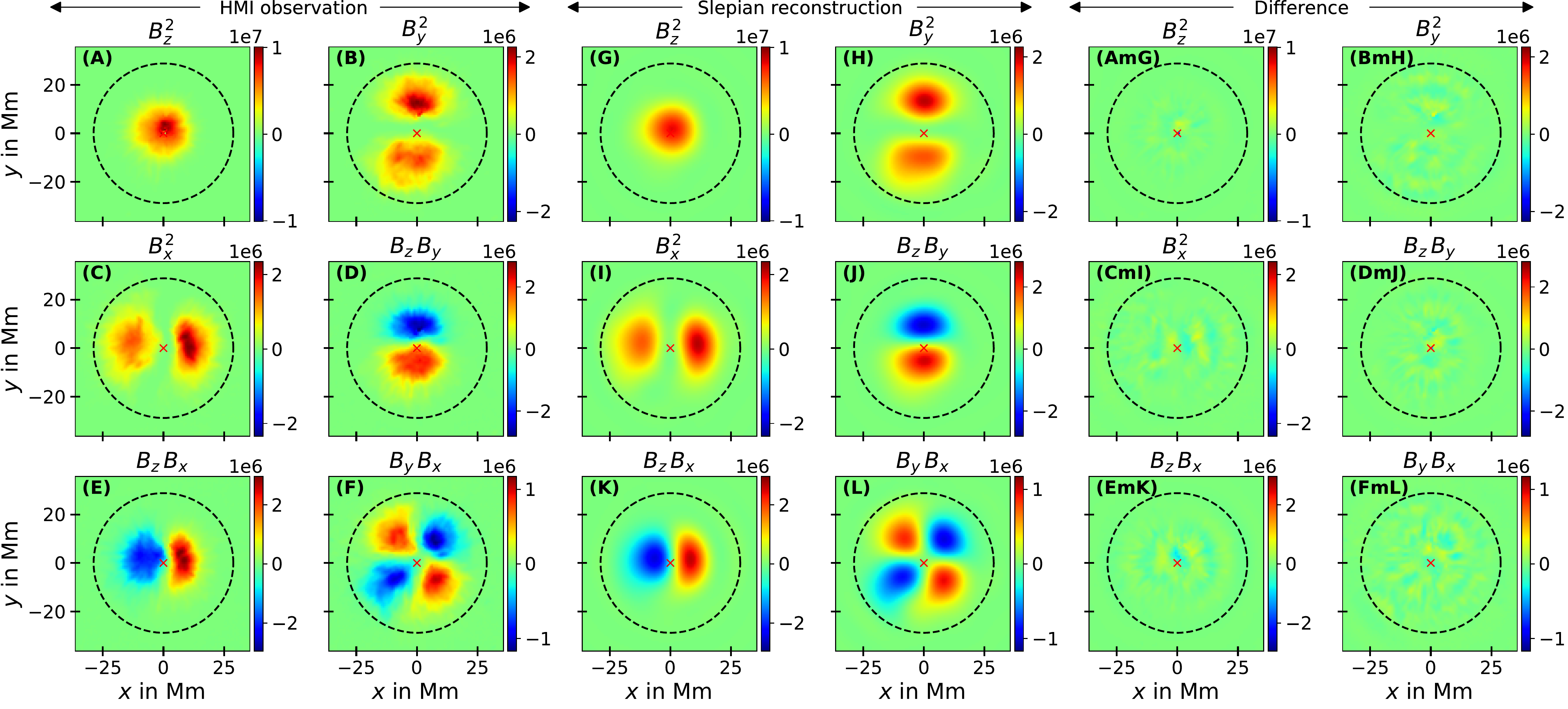}\par\medskip
  \includegraphics[width=\textwidth]{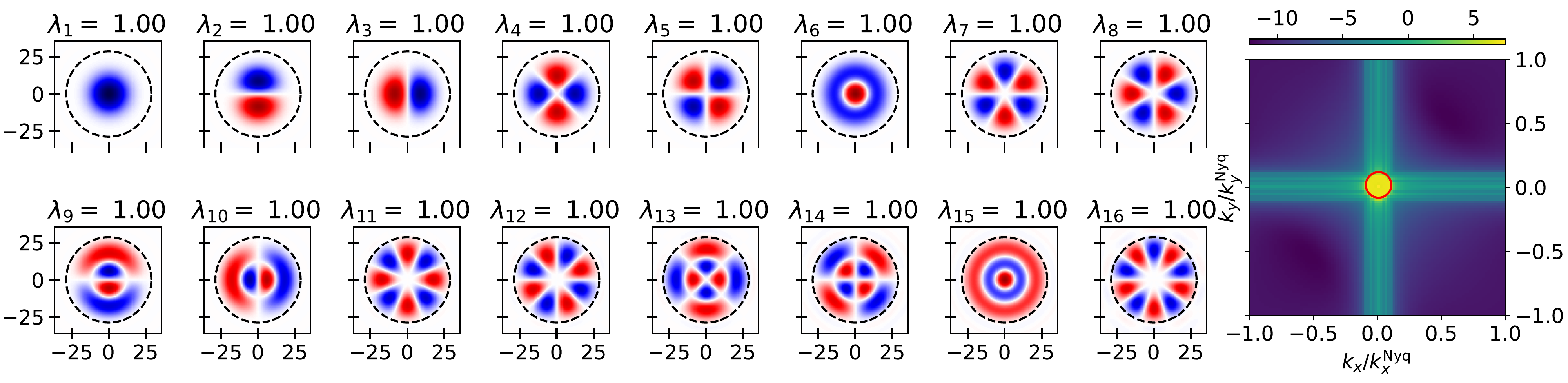}\par\medskip        
\caption{Sunspot from NOAA 11084/HARP 71 at 00:00:00 of 2nd July, 2010. Panels~(A)-(F) show the six independent components of Lorentz-stress computed using HMI vector magnetograms. Panels~(G)-(L) show the corresponding reconstructed maps using 38 Slepian functions. Panel (AmG), following the nomenclature ``A~minus~G", represents the difference between panel~(A) and panel~(G). Similarly, (BmH), (CmI), and so on, represent corresponding differences according to this nomenclature. The two lower panels show the first 16 basis functions and the concentration of energy in the spectral domain given by $\sum_{\alpha=1}^{38} \lambda_{\alpha} \, G^2_{\alpha}(\bq)$, in logarithmic scale. These basis functions' subplots share the same $x, y$ axes in units of Mm as those in panels~(A)-(F). The \textit{black dashed} circle represents the contour $\mathcal{R}$ in the physical space and the spectral space contour $\mathcal{Q}$ is denoted by the \textit{red} circle.}
\label{fig:Sunspot}
\end{figure}

\begin{figure}
\centering
  \includegraphics[width=\textwidth]{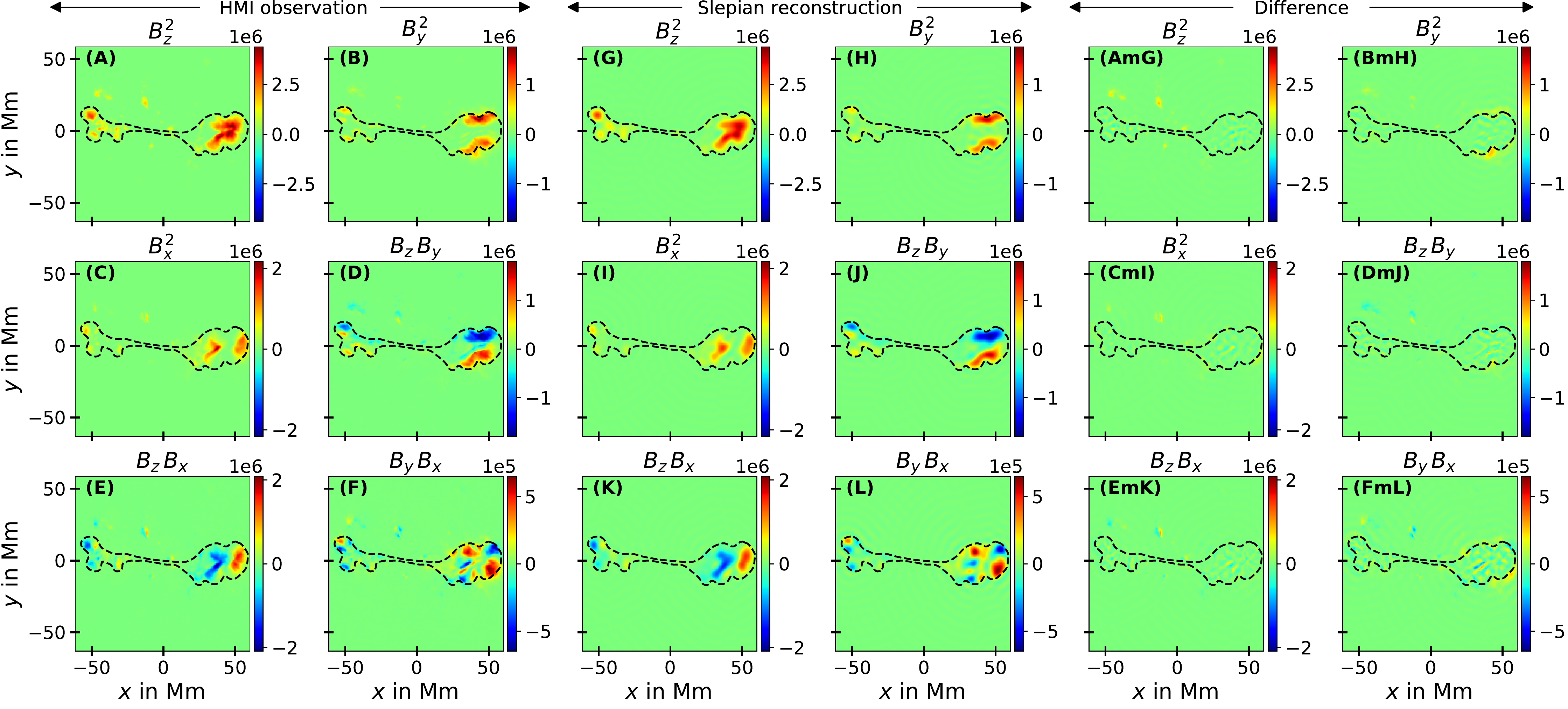}\par\medskip
  \includegraphics[width=\textwidth]{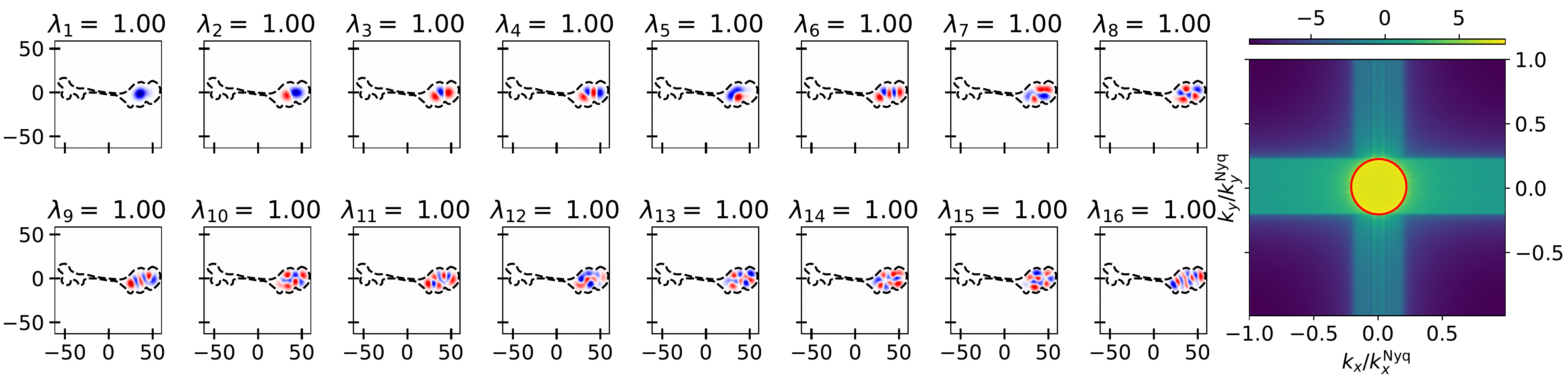}\par\medskip        
\caption{Active region from NOAA 12757/HARP 7405 averaged over two 720s HMI measurements from 15:00:00 - 15:24:00 of 26th January, 2020. Panels~(A)-(F) show the six independent components of Lorentz-stress computed using HMI vector magnetograms. Panels~(G)-(L) show the corresponding reconstructed maps using 119 Slepian functions. As per the nomenclature adopted in Fig.~(\ref{fig:Sunspot}), the panels showing the differences between HMI and reconstructed maps are named as (AmG), (BmH), (CmI), and so on. The two lower panels show the first 16 basis functions and the concentration of energy in the spectral domain given by $\sum_{\alpha=1}^{119} \lambda_{\alpha} \, G^2_{\alpha}(\bq)$, in logarithmic scale. These basis functions' subplots share the same $x, y$ axes in units of Mm as those in panels (A)-(F). The physical space contour $\mathcal{R}$ is denoted by the closed \textit{black dashed} curve around the active region and the spectral space contour $\mathcal{Q}$ is denoted by the \textit{red} circle.}
\label{fig:AR11082}
\end{figure}

Inspired from the solar surface observations, the existence of a near-surface 3D perturbation $p(\bx,z)$ may be confined to a spatial patch $\mathcal{R}$ and a spectral patch $\mathcal{Q}$. As shown in \cite{Simons&Wang_2011} which is henceforth referred to as SW11, an orthogonal family of functions $g_{\alpha}(\bx)$ which are optimally concentrated within a closed region $\mathcal{R}$ in the spatial domain, but strictly bandlimited within a closed region $\mathcal{Q}$ in Fourier domain, may be defined. Consequently, the perturbation $p(\bx,z)$ may be expressed in the corresponding Slepian basis as
\begin{equation} \label{eqn: slepian_decomposition}
    p(\bx,z) = \sum_{\alpha} p_{\alpha}(z) \, g_{\alpha}(\bx) \,.
\end{equation}
It should be noted that although, for simplicity, we choose a scalar perturbation $p$ above, the same treatment may be applied to vectorial or tensorial perturbation as well, such as flows and Lorentz-stresses, respectively. Since the current study is devoted to inferring the near-surface magnetization, Figs.~\ref{fig:Sunspot} and \ref{fig:AR11082} demonstrate the decomposition and subsequent reconstruction of the six independent components of surface Lorentz-stress using Slepian basis functions. For both the figures, panels (A)-(F) show the six independent components of the Lorentz-stress tensor $\cH$ at the solar surface $(z=0)$, calculated using $(B_x, B_y, B_z)$ from the observed HMI vector magnetograms. Using the orthonormality relation given in Eqn.~(\ref{eq: ortho_slepian}), the coefficients $p_{\alpha}$ for each of these six components is found by decomposing the maps in panels (A)-(F) in the basis of Slepian functions $g(\bx)$ as follows 
\begin{equation}
    p_{\alpha}(z=0) = \int_{\mathbb{R}^2} p(\bx, z=0) \, g_{\alpha}(\bx) \rmd \bx \, .
\end{equation}
The number of basis functions used to decompose each of the Lorentz-stress components is given by the Shannon number N2D (see Eqn.~[\ref{eq: shannon_number_eq}]). Panels (G)-(L) show the reconstructed $\cH$ components corresponding to (A)-(F). The objective of this demonstration is to show that an optimally chosen Slepian basis can be used to efficiently represent the desired structure in the domain of interest given by the \textit{black dashed} line $\mathcal{R}$. It should be mentioned here that the contour $\mathcal{R}$, plotted as \textit{black dashed} lines, are constructed by, first, hand-picking points around the region of interest inspired from the HMI maps in panels (A)-(F) and subsequently, a smooth curve is constructed by connecting the chosen points via smooth cubic splines using an interpolating function in the \texttt{scipy} routine. The difference between panels (A)-(F) and its respective reconstructed map in panels (G)-(L) is used to demonstrate the accuracy of reconstruction in panels (AmG)-(FmL). The nomenclature chosen for the difference maps is as follows: the difference map between panel (A) and panel (G) is named (AmG), which stands for ``A minus G". The same naming rule is used to make the panels (BmH)-(FmL).

The last two rows in Figs.~\ref{fig:Sunspot} and \ref{fig:AR11082} show the first 16 basis functions arranged in descending order of energy concentration within $\mathcal{R}$. The last panel shows the concentration of energy within the spectral contour $\mathcal{Q}$ marked by a \textit{red circle}. The colormap indicates the quantity $\sum_{\alpha=1}^{N2D} \lambda_{\alpha} \, G^2_{\alpha}(\bq)$, plotted in logarithmic scale. The bright \textit{yellow} illuminated patch within $\mathcal{Q}$ shows that the Slepian basis functions are strictly bandlimited. In these demonstrative figures, $\mathcal{Q}$ is chosen to be a circle in spectral space. The value of radius $|q|$ is chosen by visual inspection of the 2D fourier transform of panels (A)-(F), so as to span the region of maximal power in spectral space. An increase in $|q|$ comes with an increase in the number of Slepian functions as larger number of basis functions is needed to resolve finer and finer structures within $\mathcal{R}$.

The chosen sunspot is obtained from 720s HMI vector magnetogram measurement for NOAA 11084 or HARP 71 with the time-stamp at 00:00:00 of 2nd July, 2010\footnote[1]{JSOC links to download the vector magnetogram used for the sunspot NOAA 11084 or HARP 71: \href{http://jsoc.stanford.edu/SUM97/D1038123917/S00000/Br.fits}{$B_r$}, \href{http://jsoc.stanford.edu/SUM97/D1038123917/S00000/Bt.fits}{$B_{\theta}$}, \href{http://jsoc.stanford.edu/SUM97/D1038123917/S00000/Bp.fits}{$B_{\phi}$}.}. The first two rows in Figs.~\ref{fig:Sunspot}(A)-(F) on the left panel, show the components of $\bfB \bfB$ constructed from the observed vector magnetogram while Figs.~\ref{fig:Sunspot}(G)-(L), on the right panel, show the corresponding reconstructed maps from the Slepian decomposition as in Eqn.~(\ref{eqn: slepian_decomposition}). The panels below show the first 16 basis functions (out of a total of 38 used) arranged according to decreasing order in $\lambda_{\alpha}$, given by
\begin{equation}
    \lambda_{\alpha} = \frac{\int_{\mathcal{R}} g^2(\bx) \, \rmd \bx}{\int_{\mathbb{R}} g^2(\bx) \, \rmd \bx} \, ,
\end{equation}
which indicates the concentration of the basis function withing the domain of interest $\mathcal{R}$ in physical space. So, $\lambda_{\alpha}=1$ indicates that the basis functions $g_{\alpha}(\bx)$ is entirely contained within $\mathcal{R}$ while $\lambda_{\alpha} < 1$ indicates leakage of the basis function energy beyond the boundary of $\mathcal{R}$. The effective number of basis functions to be used is determined by the Shannon number N2D given by the sum of all $\lambda_{\alpha}$. For the shown circular region $\mathcal{R}$ in physical space and $\mathcal{Q}$ in spectral space, N2D $\approx 38$. As mentioned in Appendix~\ref{sec: slepian_functions_appendix}, N2D depends on the extent of the contour $\mathcal{Q}.$ In this example, the radius of the circle $|q|$ in spectral space is chosen by visual inspection to capture all but the extremely small-scale features such as the tiny fibril-like structures in the sunspot.

Fig.~\ref{fig:AR11082} demonstrates the decomposition of Lorentz-stress constructed using HMI vector magnetograms obtained for an active region NOAA 12757 or HARP 7405. The plotted Lorentz-stress components were obtained after averaging over two $\bfB \bfB$ obtained using 720s HMI measurements from time-stamp 15:00:00 to 15:24:00 on 26th January, 2020\footnote[2]{JSOC links to download the first 720s vector magnetogram used for the active region NOAA 12757 or HARP 7405: \href{http://jsoc.stanford.edu/SUM41/D1265819836/S00000/Br.fits}{$B_r$}, \href{http://jsoc.stanford.edu/SUM41/D1265819836/S00000/Bt.fits}{$B_{\theta}$}, \href{http://jsoc.stanford.edu/SUM41/D1265819836/S00000/Bp.fits}{$B_{\phi}$}. \\
JSOC links to download the second 720s vector magnetogram used for the active region NOAA 12757 or HARP 7405: \href{http://jsoc.stanford.edu/SUM67/D1265819940/S00000/Br.fits}{$B_r$}, \href{http://jsoc.stanford.edu/SUM67/D1265819940/S00000/Bt.fits}{$B_{\theta}$}, \href{http://jsoc.stanford.edu/SUM67/D1265819940/S00000/Bp.fits}{$B_{\phi}$}.}. This demonstration is supposed to be an extreme case of use of Slepian functions where a relatively large number of basis functions are used to capture the smallest detail. For the chosen contours in physical space $\mathcal{R}$ and spectral space $\mathcal{Q}$, the Shannon number N2D $\approx 119$. This significantly larger Shannon number as compared to the sunspot example in Fig~\ref{fig:Sunspot} is because this is an extreme demonstration where even the very fine structure is resolved by the Slepian basis. In practice, when carrying out inversions, similar choices on the maximum wavenumber $|q|$ may be made to capture broader features and ignore the very fine structure. This would require fewer Slepian functions as compared to this extreme demonstration.
Once the perturbation $p(\bx,z)$ is decomposed in the Slepian basis, the coupling matrix $\Lambda^p_{\bk\bk'}$ maybe be written as
\begin{equation}
    \Lambda^p_{\bk\bk'} = \int \rmd z \, p_{\alpha}(z) \, \cK^{(\mathrm{cart})}_{\bk\bk'}(z) \int_{\mathcal{R}} g_{\alpha}(\bx) \, e^{i (\bk - \bk') \cdot \bx} \, \rmd \bx \, ,
\end{equation}
where $\cK_{\bk\bk'}^{(\mathrm{cart})}$ are the kernels in traditional Cartesian mode-coupling where plane-wave basis functions are used. For instance, in the case where the perturbation is the second-rank tensor $\bfB\bfB$, $\cK_{\bk\bk'}^{(\mathrm{cart})}$ is the tensorial kernel given by $\boldsymbol{\cK}_{\bk\bk'}$ as in Eqn.~(\ref{eq: kernel_comps}). Further, if $G_{\alpha}(\bq)$ is the bandlimited representation of the spatially concentrated Slepian functions $g_{\alpha}(\bx)$ then,
\begin{equation}
    g_{\alpha}(\bx) = (2\pi)^{-2} \, \int_{\mathcal{Q}} G_{\alpha}(\bq) \, e^{i \bq \cdot \bx} \, \rmd \bq \, ,
\end{equation}
where $\mathcal{Q}$ is the closed curve in the spectral space that encloses all the spectral wavenumbers  of interest, $\bq$. The coupling matrix then takes the form
\begin{equation} \label{eqn:put_slep_in_Lambda}
    \Lambda_{\bk\bk'}^p = (2\pi)^{-2} \, \int \rmd z \, p_{\alpha}(z) \, \cK_{\bk\bk'}^{(\mathrm{cart})}(z) \int_{\mathcal{Q}} \int_{\mathcal{R}} G_\alpha(\bq) \, e^{i (\bq +\bk-\bk')\cdot\bx} \, \rmd\bx \, \rmd\bq \, .
\end{equation}
Defining $\bk' - \bk = \Delta \bk$ and subsequently using the following identity of the Cartesian Slepian basis from Eqn.~(37) in SW11,
\begin{equation}
    (2\pi)^{-2} \, \int_\mathcal{Q} \int_\mathcal{R}e^{i (\bq- \Delta \bk) \cdot \bx} \, \rmd \bx \, G_{\alpha}(\bq) \, \rmd \bq = \lambda_{\alpha} \, G_{\alpha}(\Delta \bk) \, , \qquad \Delta \bk \in \mathcal{Q} \, ,
\end{equation}
and discretizing the depth $z$ in the basis of B-splines $f_j(z)$, the coupling matrix expression simplifies to
\begin{eqnarray} 
    &&\Lambda^p_{\bk \bk'} =  \sum_j \, p_{j\alpha} \, \cK_{\bk \bk'}^{j\alpha \, (\mathrm{slep})} \, , \label{eqn: slepian_inverse_problem1}\\
    &&\mathrm{where} \quad \cK_{\bk \bk'}^{j\alpha \, (\mathrm{slep})} = \cK_{\bk \bk'}^{j \, (\mathrm{cart})} \, \lambda_{\alpha} \, G_{\alpha}(\bk' - \bk) \, , \\
    &&\mathrm{and} \quad \cK_{\bk \bk'}^{j \, (\mathrm{cart})} = \int f^j(z) \, \cK^{(\mathrm{cart})}_{\bk\bk'}(z) \rmd z \, . \label{eqn: slepian_inverse_problem2}
\end{eqnarray}
Therefore, the Cartesian plane-wave kernels can be seamlessly converted to the Slepian basis kernels via a linear transformation. For the perturbation $p_{j\alpha}$, information of the vertical profile is captured by the index $j$ while the distribution in the horizontal plane is captured by the index $\alpha$ labeling the Slepian basis functions. Therefore, there is no $\bk$ or $\bk'$ dependence in the way the perturbation is modelled. This results in all the modes in the coupling matrix $\Lambda_{\bk\bk'}^p$ in the mode superset $\mathbf{K}_{\mathrm{super}}$ together constraining each of the model parameters $p_{j\alpha}$. Compared to Eqn.~(\ref{eq: coupling_matrix_BB}), this drastically increases the confidence in inferring the model parameters by increasing the signal-to-noise ratio as a result of enhancing the number of data constraints.

Expanding on the above, it may be instructive to consider an example where $\mathcal{Q}$ is a circle and there are $M$ gridpoints in $q_x$ within the circle. Since $g(\bx)$ is real, this imposes the symmetry $G(\bq) = G^*(-\bq)$. Consequently, there are $\mathcal{O}(M^2/2)$ combinations of $\bq$ that carry unique information within $\mathcal{Q}$. In the case of the plane-wave basis, the inversion for each of these $\mathcal{O}(M^2/2)$ different $\bq'$s was constrained by its own separate set of $\bk, \bk'$. However, when using Eqn.~(\ref{eqn: slepian_inverse_problem1}) for inversion, the data constraints are combined. So, even if the number of basis functions N2D is comparable to the number of $(q_x, q_y)$ pairs used for the plane-wave basis, the Slepian approach increases the number of constraints for each unknown by a factor of $\mathcal{O}(M^2/2)$ times. Consequently, the signal-to-noise ratio of each unknown drops by a factor of around $M/\sqrt{2}$.

\subsection{Non-linear inversion: With a divergence-free $\bfB$ as the model parameter} \label{sec: inversion_B}
This section is devoted to setting up the inversion formalism when trying to infer the magnetic field $\bfB$ directly instead of the Lorentz-stress. Since $\bnabla \cdot \bfB = 0$, the magnetic field can be written in the solenoidal form
\begin{equation} \label{eq: B_solenoidal}
    \bfB = \bnabla \times \left(\bnabla \times \left(P \ev{z} \right) \right) + \bnabla \times \left(T \ev{z} \right) \, ,
\end{equation}
where $P$ and $T$ are the scalar stream-functions that entirely define a three-dimensional solenoidal field. As for the components of Lorentz-stress tensor in Section~(\ref{sec: BB_inversion}), we can decompose the scalars $P$ and $T$ as
\begin{equation}
    P = \sum_j \int f^j \, P_{j\bq} \, e^{i \bq \cdot \bx} \, \rmd \bq \, , \qquad T = \sum_j \int f^j \, T_{j \bq} \, e^{i \bq \cdot \bx} \, \rmd \bq \, .
\end{equation}
Now, using the vector calculus identity $\bnabla \times \left( \bnabla \times \bfA \right) = \bnabla \left( \bnabla \cdot \bfA \right) - \nabla^2 \bfA$, it can be shown that the poloidal and toroidal term reduces to
\begin{eqnarray}
    \bnabla \times \left(\bnabla \times \left(P \ev{z} \right) \right) &=& \int \left(\ev{z} \, q^2 \, f^j + i \bq \dot{f}^j \right) \, P_{j \bq} \, e^{i \bq \cdot \bx} \, \rmd \bq \, ,\label{eq:P}\\ 
    \bnabla \times \left( T \ev{z}\right) &=& - i \int (\ev{z} \times \bq) \, f^j \, T_{j \bq} \, e^{i \bq \cdot \bx} \, \rmd \bq \, , \label{eq:T}
\end{eqnarray}
where $\dot{f}^j = \rmd f^j/\rmd z$ and Einstein's summation convention over $j$ is assumed. Substituting Eqn.~(\ref{eq:P}) and Eqn.~(\ref{eq:T}) into Eqn.~(\ref{eq: B_solenoidal}), we get
\begin{equation}
    \bfB = \int \left(\cP_{j \bq} \, P_{j \bq} + \cT_{j \bq} \, T_{j \bq} \right) \, e^{i \bq \cdot \bx} \rmd \bq \, ,
\end{equation}
where we define vectors $\cP_{j \bq} = \ev{z} \, q^2 \, f^j + i \bq \dot{f}^j $ and $\cT_{j \bq} = i\,(\bq \times \ev{z}) \, f^j$ for convenience. Consequently, the Lorentz-stress $\bfB \bfB$ takes the form
\begin{equation} \label{eq: BB_from_solenoidal_B}
    \bfB \bfB = \int \int \left( \cP_{j \bq} \, \cP_{j' \bq'} \, P_{j \bq} \, P_{j' \bq'} +  \cT_{j \bq} \, \cT_{j' \bq'} \, T_{j \bq} \, T_{j' \bq'} + \cP_{j \bq} \, \cT_{j' \bq'} \, P_{j \bq} \, T_{j' \bq'} + \cT_{j \bq} \, \cP_{j' \bq'} \, T_{j \bq} \, P_{j' \bq'}\right) e^{i(\bq + \bq') \cdot \bx} \, \rmd \bq \, \rmd \bq' \, .
\end{equation}
It may be noted that when compared to Eqn.~(\ref{eqn:put_slep_in_Lambda}), it can be inferred immediately that decomposing such a perturbation in the Slepian basis would not yield the desired simplification due to the presence of the double integral over $\bq$ and $\bq'$. Consequently, the calculations are restricted to the plane-wave basis in the horizontal plane. Substituting the expression of $\bfB \bfB$ from Eqn.~(\ref{eq: BB_from_solenoidal_B}) in Eqn.~(\ref{eq: coupling_matrix}), the resultant coupling matrix can be expressed as
\begin{equation} \label{eq: coupling_matrix_B}
\begin{split}
    \Lambda_{\bk\bk'}^{\mathrm{mag}} = \int \int & \left[\left(\cP_{j\bq} \, \cP_{j'\bq'} : \boldsymbol{\cK}_{\bk\bk'} \right) \, P_{j \bq}\,P_{j' \bq'} + \left(\cT_{j\bq} \, \cT_{j'\bq'} : \boldsymbol{\cK}_{\bk\bk'} \right) \, T_{j \bq}\,T_{j' \bq'} \right. \\ 
    &+ \left. \left(\cP_{j\bq} \, \cT_{j'\bq'} : \boldsymbol{\cK}_{\bk\bk'} \right)\, P_{j \bq}\,T_{j' \bq'} + \left(\cT_{j\bq} \, \cP_{j'\bq'} : \boldsymbol{\cK}_{\bk\bk'} \right)\, T_{j \bq}\,P_{j' \bq'}\right] \, \rmd z \, \rmd \bq \, ,
\end{split}
\end{equation}
where the selection rule $\bk' = \bk + \bq + \bq'$ is imposed due to the integral $\int \exp{\left[i(\bk + \bq + \bq' - \bk') \cdot \bx\right]} \, \rmd^2 \bx$. Eqn.~(\ref{eq: coupling_matrix_B}) shows that a non-linear inverse problem is necessary to find the coefficients $P_{j\bq}$ and $T_{j\bq}$. The terms involving tensor contractions in the parenthesis are the ``effective kernels" for inversion. The following definitions are used to find the explicit expressions for these effective kernels.
\begin{eqnarray}
    \cP_{j \bq} \cdot \ev{z} &=& q^2 \, f^j = \alpha_{q}^j \, ,\\
    \cP_{j \bq} \cdot \ev{x} &=& i \, \bq \cdot \ev{x} \, \dot{f}^j = \beta_{\bq \bx}^j \, ,\\
    \cP_{j \bq} \cdot \ev{y} &=& i \, \bq \cdot \ev{y} \, \dot{f}^j = \beta_{\bq \by}^j \, ,\\
    \cP_{j \bq} \cdot \ev{k} &=& i \, \bq \cdot \ev{k} \, \dot{f}^j = \beta_{\bq \bk}^j \, ,\\
    \cT_{j \bq} \cdot \ev{z} &=& 0 \, ,\\
    \cT_{j \bq} \cdot \ev{x} &=& i \ev{x} \cdot (\ev{q} \times \ev{z}) \, q \, f^j = i \, \ev{z} \cdot (\ev{x} \times \ev{q}) \, q \, f^j = \gamma^j_{\bq \bx} \, ,\\ 
    \cT_{j \bq} \cdot \ev{y} &=& i \ev{y} \cdot (\ev{q} \times \ev{z}) \, q \, f^j = i \, \ev{z} \cdot (\ev{y} \times \ev{q}) \, q \, f^j = \gamma^j_{\bq \by} \, ,\\
    \cT_{j \bq} \cdot \ev{k} &=& i \, \ev{k} \cdot (\ev{q} \times \ev{z}) \, q \, f^j = i \, \ev{z} \cdot (\ev{k} \times \ev{q}) \, q \, f^j = \gamma_{\bq \bk}^j \, .
\end{eqnarray}
It is to be noted that $\alpha_q^j$ depends only on the magnitude of $\bq$ and $\beta_{\bq \bk}^k, \gamma_{\bq \bk}^j$ are neither commutative nor anti-commutative in $(\bq , \bk)$. Using these, we find the effective kernels in Eqn.~(\ref{eq: coupling_matrix_B}):
\begin{eqnarray}
    \cP_{j \bq} \, \cP_{j' \bq'}:\boldsymbol{\cK}_{\bk\bk'} = &(&\beta_{\bq \bx}^j \, \beta_{\bq' \bx}^{j'} + \beta_{\bq \by}^j \, \beta_{\bq' \by}^{j'}) \, \cK_{\bk\bk'}^{xx} +  \alpha_q^{j} \, \alpha_{q'}^{j'} \, \cK_{\bk\bk'}^{zz} + \beta_{\bq \bk}^{j} \, \alpha_{q'}^{j'} \, \cK_{\bk\bk'}^{zk} + \beta_{\bq \bk'}^{j} \, \alpha_{q'}^{j'} \, \cK_{\bk\bk'}^{zk'} + \alpha_q^{j} \, \beta_{\bq' \bk}^{j'} \, \cK_{\bk\bk'}^{kz} \\ \nonumber
    &+& \alpha_q^{j} \,\beta_{\bq' \bk'}^{j'} \, \cK_{\bk\bk'}^{k'z} + \beta_{\bq \bk}^{j} \, \beta_{\bq' \bk}^{j'} \, \cK_{\bk\bk'}^{kk} + \beta_{\bq\bk'}^{j} \, \beta_{\bq'\bk'}^{j'} \, \cK_{\bk\bk'}^{k'k'} +   \beta_{\bq\bk}^{j} \, \beta_{\bq'\bk'}^{j'} \, \cK_{\bk\bk'}^{k'k} +  \beta_{\bq\bk'}^{j} \, \beta_{\bq'\bk}^{j'} \, \cK_{\bk\bk'}^{kk'} \, ,\\
    \cT_{j \bq} \, \cT_{j' \bq'} : \boldsymbol{\cK}_{\bk\bk'} = &(&\gamma_{\bq \bx}^{j} \, \gamma_{\bq' \bx}^{j'} + \gamma_{\bq \by}^{j} \, \gamma_{\bq' \by}^{j'}) \, \cK_{\bk\bk'}^{xx} + \gamma_{\bq \bk}^{j} \, \gamma_{\bq' \bk}^{j'} \, \cK_{\bk\bk'}^{kk} + \gamma_{\bq \bk'}^{j} \, \gamma_{\bq' \bk'}^{j'} \, \cK_{\bk\bk'}^{k'k'} + \gamma_{\bq \bk'}^{j} \, \gamma_{\bq' \bk}^{j'} \, \cK_{\bk\bk'}^{kk'} \\ \nonumber
    &+& \gamma_{\bq \bk}^{j} \, \gamma_{\bq' \bk'}^{j'} \, \cK_{\bk\bk'}^{k'k} \, ,\\
    \cP_{j \bq} \, \cT_{j' \bq'} : \boldsymbol{\cK}_{\bk\bk'} = &(&\beta_{\bq \bx}^j \, \gamma_{\bq' \bx}^{j'} + \beta_{\bq \by}^{j} \, \gamma_{\bq' \by}^{j'}) \, \cK_{\bk\bk'}^{xx} +\alpha_q^j \, \gamma_{\bq' \bk}^{j'} \, \cK_{\bk\bk'}^{kz} + \alpha_q^j \, \gamma_{\bq' \bk'}^{j'} \, \cK_{\bk\bk'}^{k'z} + \beta_{\bq \bk}^j \, \gamma_{\bq' \bk}^{j'} \, \cK_{\bk\bk'}^{kk} + \beta_{\bq \bk'}^j \, \gamma_{\bq' \bk'}^{j'} \, \cK_{\bk\bk'}^{k'k'} \\ \nonumber
    &+& \beta_{\bq \bk'}^j \, \gamma_{\bq' \bk}^{j'} \, \cK_{\bk\bk'}^{kk'} + \beta_{\bq\bk}^j \, \gamma_{\bq' \bk'}^{j'} \, \cK_{\bk\bk'}^{k'k} \, ,\\
    \cT_{j \bq} \, \cP_{j' \bq'} : \boldsymbol{\cK}_{\bk\bk'} = &(&\gamma_{\bq \bx}^j \, \beta_{\bq' \by}^{j'} + \gamma_{\bq \by}^j \, \beta_{\bq' \by}^{j'}) \, \cK_{\bk\bk'}^{xx} + \gamma_{\bq \bk}^j \, \alpha_{q'}^{j'} \, \cK_{\bk\bk'}^{zk} + \gamma_{\bq \bk'}^j \, \alpha_{q'}^{j'} \, \cK_{\bk\bk'}^{zk'} + \gamma_{\bq \bk}^j \, \beta_{\bq'\bk}^{j'} \, \cK_{\bk\bk'}^{kk} + \gamma_{\bq \bk'}^j \, \beta_{\bq'\bk'}^{j'} \, \cK_{\bk\bk'}^{k'k'} \\ \nonumber
    &+& \gamma_{\bq \bk'}^j \, \beta_{\bq'\bk}^{j'} \, \cK_{\bk\bk'}^{kk'} + \gamma_{\bq \bk}^j \, \beta_{\bq'\bk'}^{j'} \, \cK_{\bk\bk'}^{k'k} \, .
\end{eqnarray}

\subsection{Implication on simultaneous inference of sound-speed and Lorentz-stress} \label{sec: sound_speed_mag_anomaly}

In the regime where linear perturbations capture anomalies in sound-speed, the perturbation operator is $\delta \cL \bfxi = -2 \bnabla\left(\rho \, c \, \delta c \bnabla \right) \cdot \bfxi$, where $c$ and $\delta c$ are the background sound-speed and sound-speed perturbation, respectively. Assuming $\delta c = 0$ at the solar surface (which may be imposed by surface constraints from observations), it is straightforward to show that the coupling matrix is 
\begin{eqnarray}
    \Lambda_{\bk\bk'}^c &=& 2 \int_V \rho \, c \, \delta c \, \bnabla\cdot\bfxi_{\bk}\bnabla\cdot\bfxi^*_{\bk'} \, \rmd V \\
    &=& \int \rho \, \left(\dot{V}_k \, \dot{V}_{k'}^* + k\,k'\,H_{k}\,H_{k'}^* - k'\, \dot{V}_{k} \, H_{k'}^* - k \, \dot{V}_{k'}^* \, H_{k}  \right)\, (2 \, c \, \delta c_\bq) \, \rmd z \, ,
\end{eqnarray}
where $\bk' = \bk + \bq$ as in Eqn.~(\ref{eq: coupling_matrix_BB}). Therefore, sound-speed kernel is 
\begin{equation}
    \cK_{\bk\bk'}^c = \rho \left(\dot{V}_k \, \dot{V}_{k'}^* + k\,k'\,H_{k}\,H_{k'}^* - k'\, \dot{V}_{k} \, H_{k'}^* - k \, \dot{V}_{k'}^* \, H_{k}  \right) = \rho \, \cK_{\bk\bk'}^{xx} =   \rho \, \cK_{\bk\bk'}^{yy} \, .
\end{equation}
This shows that at each depth, the kernels for sound-speed anomaly $2 \, c \, \delta c$ and the strongest components of the kernels for $B_x^2$ and $B_y^2$ differ only by a multiplicative factor of $\rho$. Consequently, when carrying out a combined linear inversion for sound-speed and Lorentz-stress, the total coupling matrix is
\begin{eqnarray}
    \Lambda_{\bk\bk'} &=& \Lambda_{\bk\bk'}^{\mathrm{mag}} + \Lambda_{\bk\bk'}^c \\
    &=& \int \left[ \cK_{\bk\bk'}^{xx} \left(2 \, \rho \, c \, \delta c + B_x^2 + B_y^2 \right) + \tilde{\mathcal{B}}_{xx} \, \mathcal{H}^{xx} + \tilde{\mathcal{B}}_{yy} \, \mathcal{H}^{yy} + \mathcal{B}_{zz} \, \mathcal{H}^{zz} + \mathcal{B}_{xy} \, \mathcal{H}^{xy} + \mathcal{B}_{xz} \, \mathcal{H}^{xz} + \mathcal{B}_{yz} \, \mathcal{H}^{yz} \right] \rmd z \, .
\end{eqnarray}
Here, $\tilde{\mathcal{B}}_{xx,yy} = \mathcal{B}_{xx,yy} - \cK_{\bk\bk'}^{xx}$. Consequently, $2 \, \rho \, c \, \delta c + B_x^2 + B_y^2$ needs to be grouped as a single invertible parameter, thereby hindering an unambiguous inference of sound-speed.

\section{Discussion} \label{sec: discussion}
From its recent successes, Cartesian mode-coupling seems to be a potential candidate for complementing local helioseismic techniques. Measurements such as $B$-coefficients \citep[as defined and used in][]{woodard16, hanasoge18, Hanson21, Mani22} capture the effect of all near-surface perturbations such as flows, sound-speed and magnetic fields. Calculating sound-speed and flow kernels are straightforward due to the scalar and vectorial nature of the perturbations, respectively. As shown in D20, taking a formal approach to device an inverse problem for magnetic field poses the second rank Lorentz-stress tensor $\bfB \bfB$ as the invertible model parameter. D20 presented the form of Lorentz-stress kernels for a general configuration of $\bfB$ and found the specific expressions of these kernels in a spherical polar coordinate system useful for GMC. This study finds the explicit analytical expressions for these kernels in the Cartesian coordinate system needed in LMC.

The currently existing formalism in Cartesian mode-coupling depends on the selection-rule based inference of model parameters $P_{j\bq}$ from a subset of coupling matrices $\Lambda_{\bk\bk'}^p$ (or equivalently $B$-coefficients) for which $\bk' = \bk + \bq$ is satisfied. Moreover, the decomposition of arbitrary shaped perturbations (such as circular for a sunspot or an averaged supergranule) in horizontal plane in the basis of $e^{i \bk \cdot \bx}$, increases the number of model parameters needed to resolve the perturbation. Therefore, the currently existing Cartesian mode-coupling formalism suffers from a compromise in the number of data constrains which is further exacerbated by the need for a large number of model parameters. Both of this ultimately makes the inverse problem poorly conditioned and consequently reduces the precision of inversion. This paper prescribes the method of using Slepian functions as basis elements in the horizontal plane. Mostly used in terrestrial and planetary sciences, Slepian functions are specially designed to build optimal basis for an arbitrary patch in physical and spectral domains (chosen by respecting the shape of perturbations in the respective domains). Therefore, for a circular perturbation such as a sunspot or an averaged supergranule, the basis elements respect the circular geometry (as shown in Fig.~\ref{fig:Sunspot}) and minimize the number of basis functions needed. The inference of all the model parameters happen at once since they do not have a wavenumber dependence as seen in Eqns.~(\ref{eqn: slepian_inverse_problem1})-(\ref{eqn: slepian_inverse_problem2}). Instead, the model parameters depend only on the knot location $j$ in depth and Slepian index $\alpha$. Once again, taking the simple example of a circular perturbation, it is easy to interpret the combined constraining of all data $\Lambda_{\bk\bk'}$ for all the Slepian parameters. Since the coupling matrix is formed from the cross-correlation of the mode-amplitude of plane waves propagating in the medium, it would take all such plane waves (in principle infinite) to constrain a structure which is circular. From this intuitive argument, it is readily seen why choosing a plane-wave basis is sub-optimal for an arbitrarily-shaped perturbation and only those functions that respects the shape of the perturbation would constitute an optimal basis. Finally, for a plane-wave basis expansion, the misfit functions (when carrying out the inversion), take in account features outside the region of interest $\mathcal{R}$ in the data-cube. Depending on whether or not the contribution from the region outside $\mathcal{R}$ is substantial, the misfit function may or may not suffer from data-contamination. Using the Slepian functions which are space-limited within $\mathcal{R}$ would render the misfit functions sensitive solely to the observations within the region of interest, thereby eliminating this data-contamination from the rest of the unwanted pixels in the data-cube.

Formalism for inverting for both, (a) the six independent components of the Lorentz-stress tensor in a Slepian basis, as well as (b) the two independent scalar stream functions $(P, \, T)$ for a solenoidal magnetic field in a plane-wave basis, are presented. Eqn.~(\ref{eq: coupling_matrix_BB}) shows that inferring Lorentz-stress components is achievable through a linear inverse problem while Eqn.~(\ref{eq: coupling_matrix_B}) shows that a non-linear treatment is needed to infer the solenoidal magnetic field. In Eqn.~(\ref{eq: BB_from_solenoidal_B}), the presence of a double integral on horizontal wavenumbers $\bq$ and $\bq'$, prohibit the use of Slepian functions for the case of a solenoidal magnetic field. For regions with sufficiently strong magnetic fields, such as active regions and sunspots, vector magnetogram observations could be used as a surface constraint. If $f_N$ refers to the B-spline that provides local support at $R_{\odot}$, the Lorentz-stress components $\mathcal{H}^{\gamma \delta}_{N \bq}$ (or its equivalent decomposition in the Slepian basis), or the magnetic field stream function coefficients $P_{N \bq}$ and $T_{N \bq}$ can be calculated from vector magnetograms and held fixed during inversions. Such constraints may be reliably applied for active regions and sunspots where the field strength is of the order of a few hundred Gauss. \cite{Hoeksema14} discuss disambiguation techniques for strong and noisy magnetic field regions in the Sun. These disambiguation methods are a part of the \href{http://jsoc.stanford.edu/}{JSOC} pipeline for processing vector magnetograms.

Care needs to be taken when analyzing quiet Sun patches with weak magnetic fields where disambiguation techniques are not reliable. Traditional averaging of $b$-coefficients to increase their signal-to-noise ratio would still produce accurate line-of-sight component $B_z$, but could fail to produce accurate measures of surface $B_x$ and $B_y$. Fortunately, Section~(\ref{sec: BB_inversion}) shows that $B_z^2$ is one of the components of Lorentz-stress to which helioseismic observations are the most sensitive. Consequently, $B_z$ could still be used as a constraint for weak magnetic field patches on the Sun. 

\section{Acknowledgement}
This work was supported, in part by the Elisabeth~H. and F.~A.~Dahlen~Fund award by the Department of Geosciences, Princeton University. The author would like to thank Prasad Mani, Samarth G. Kashyap, Prof.~Shravan M. Hanasoge of TIFR, Mumbai, Dr.~Christopher Hanson of NYU, Abu Dhabi and Prof.~Jeroen Tromp of Princeton University, USA for numerous discussions and valuable feedback on the manuscript. The author especially acknowledges innumerable disucssions with Prof. Frederik J. Simons of Princeton University on Slepian functions and his extensive \href{https://geoweb.princeton.edu/people/simons/software.html}{software toolkit} on Slepian functions.

\appendix

\section{Decomposing LCT flow-maps in Slepian basis} \label{sec: AppendixA}
\begin{figure}
\centering
  \includegraphics[width=\textwidth]{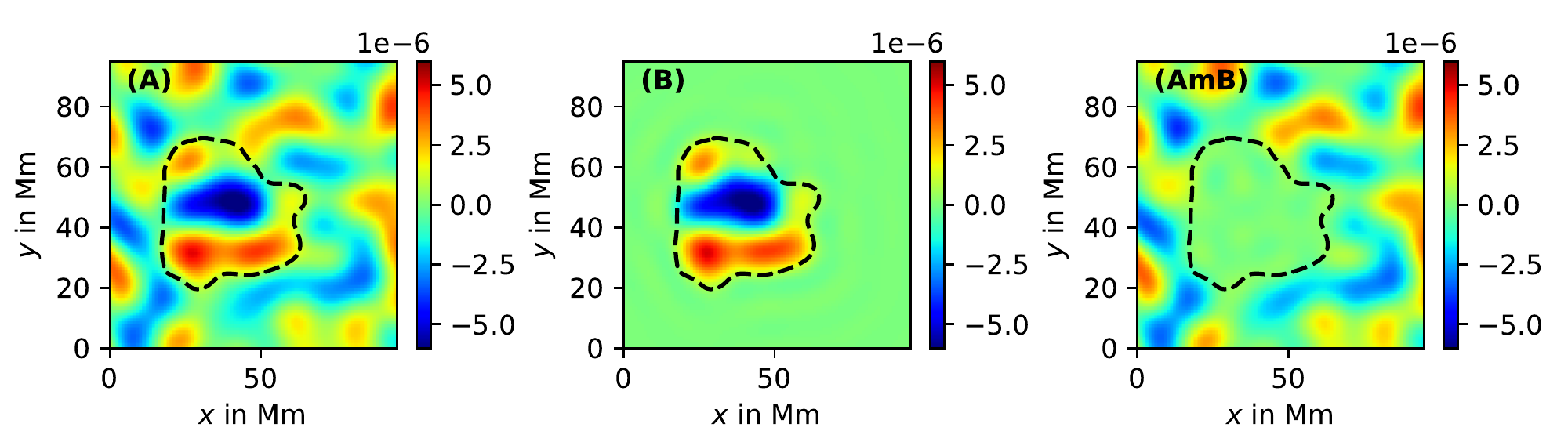}\par\medskip
  \includegraphics[width=\textwidth]{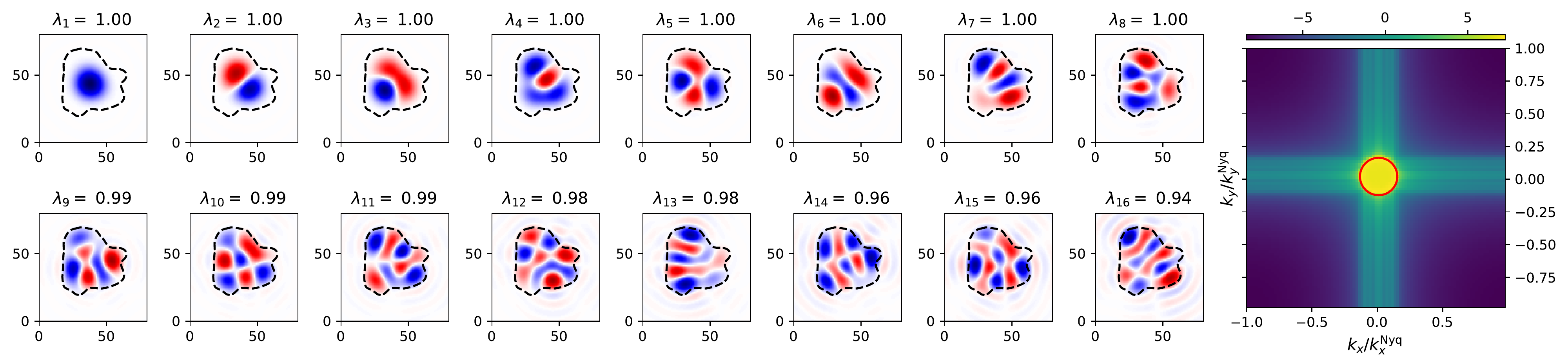}\par\medskip        
\caption{\textit{Top panel:} (A) Same as the \textit{left} panel in Fig.~6 of \cite{birch19} showing $\bnabla \cdot \bfv_h$ maps from the average of 57 emerging active regions 13.5 hours prior to emergence. (B) The corresponding reconstructed map using 27 Slepian functions. Similar to the nomenclature adopted in Fig.~(\ref{fig:Sunspot}), panel (AmB) indicates the difference between the map in (A) and that in (B). \textit{Bottom panel:} The two lower panels show the first 16 basis functions and the concentration of energy in the spectral domain given by $\sum_{\alpha=1}^{27} \lambda_{\alpha} \, G^2_{\alpha}(\bq)$, in logarithmic scale. These basis functions' subplots share the same $x, y$ axes in units of Mm as that in panel (A). The physical space contour $\mathcal{R}$ is denoted by the closed \textit{black dashed} curve and the spectral space contour $\mathcal{Q}$ is denoted by the \textit{red} circle. }
\label{fig:LCT_EAR}
\end{figure}
Even though the current study is devoted to magnetic perturbations, this section demonstrates the utility of Slepian decomposition for flow perturbations in emerging active regions. The mathematical formalism of decomposing in a Slepian basis is the same as illustrated in Section~(\ref{sec: slepian}), except that the perturbation $p(\bx)$ is the surface map of the divergence of the horizontal flow components, $\bnabla \cdot \bfv_h$. The flow-map used here is the same as that in Fig.~6 in \cite{birch19} where 57 active regions 13.5 hours prior to emergence were averaged. The \textit{black dashed} contour in Fig.~\ref{fig:LCT_EAR} shows the chosen contour in the physical domain $\mathcal{R}$. Although the exact choice of the contour, for the purpose of this paper, is to solely demonstrate the accuracy of replication of the flow-map within $\mathcal{R}$, the nearest outflow regions are included to respect anelasticity within the domain of interest. Fig.~\ref{fig:LCT_EAR}(A) shows the original flow-map, Fig.~\ref{fig:LCT_EAR}(B) shows the reconstructed map after decomposing the original flow-map into the Slepian basis and Fig.~\ref{fig:LCT_EAR}(AmB) shows the difference between panels (A) and (B). The lower two panels show the the first 16 out of a total of 27 Slepian  functions that are used to reconstruct Fig.~\ref{fig:LCT_EAR}(B) in the descending order of $\lambda_{\alpha}$. As also shown previously in Figs.~(\ref{fig:Sunspot}) and (\ref{fig:AR11082}), the last panel in the lower-right of Fig.~(\ref{fig:LCT_EAR}) shows the concentration of spectral power within the \textit{red} circle $\mathcal{Q}$. The colormap shows $\sum_{\alpha=1}^{27} \lambda_{\alpha} \, G_{\alpha}^2(\bq)$ in logarithmic scale.

\section{Slepian functions} \label{sec: slepian_functions_appendix}
\begin{figure}
    \centering
    \includegraphics[width=0.45\textwidth]{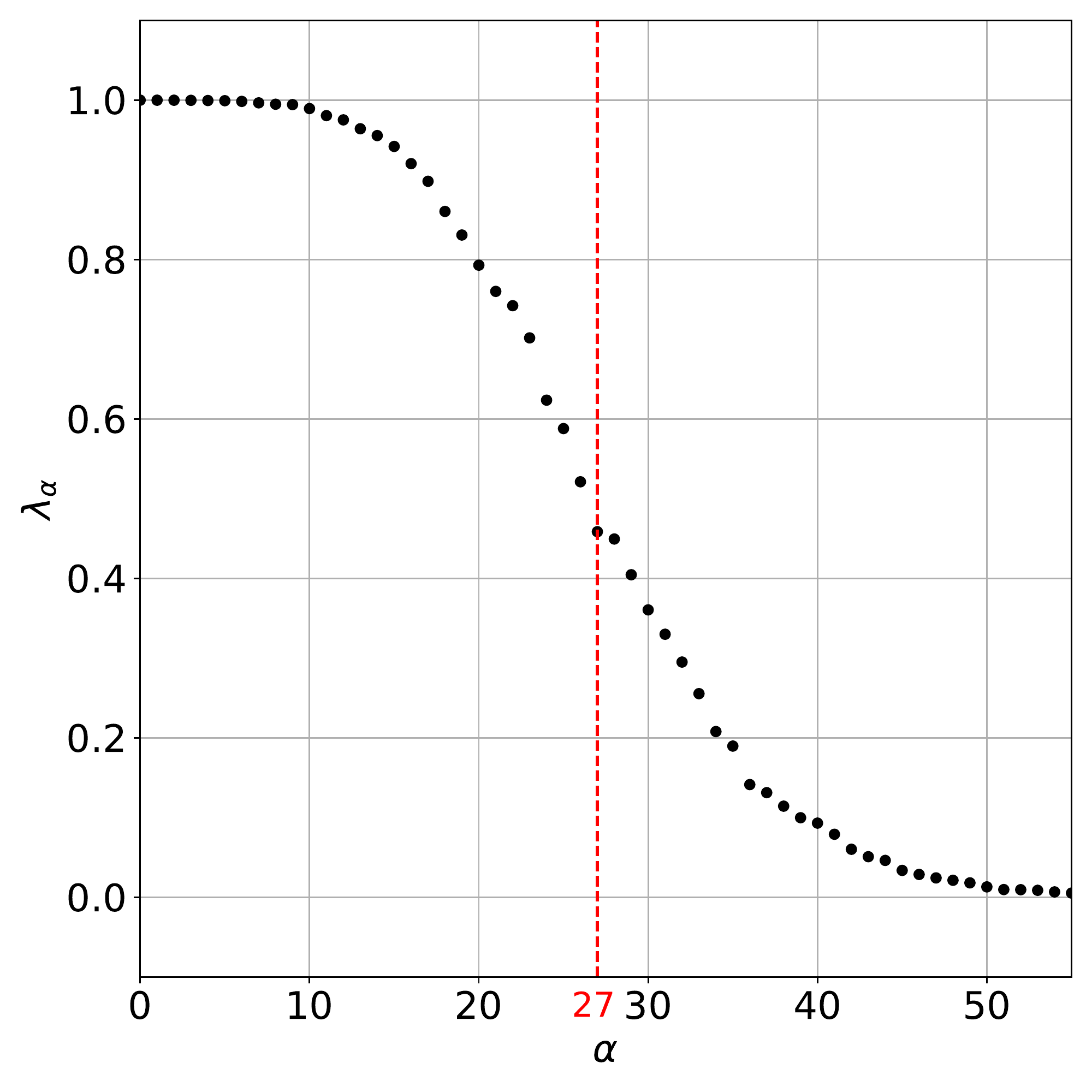}
    \caption{Plot showing the $\lambda_{\alpha}$ for the eigenfunctions using in the emerging active region demonstration in Fig.~(\ref{fig:LCT_EAR}). The red line indicates $\alpha=27$ which is used as the Shannon number N2D to calculate the effective number of eigenfunctions which are optimally concentrated within $\mathcal{R}$.}
    \label{fig:shannon_number_fig}
\end{figure}
Slepian functions have not been used in helioseismology prior to this and forms a crucial part of the method described in this study. Therefore, for the convenience of the reader, this Appendix outlines some of the basic theoretical framework of Slepian functions in a Cartesian plane. The reader is recommended to peruse SW11 for further details.

For the purpose of reference, the Slepian functions from Fig.~\ref{fig:LCT_EAR} could be used. The area of interest in space $\mathcal{R}$ is demarcated by the \textit{black dashed} line and the wavenumbers of interest are enclosed within the region $\mathcal{Q}$, shown as the \textit{red} circle. Given $\mathcal{R}$ and $\mathcal{Q}$, it is possible to define functions $g(\bx)$ which are bandlimited
\begin{equation}
    g(\bx) = (2\pi)^{-2} \int_{\mathcal{Q}} G(\bq) \, e^{i \bq \cdot \bx} \, \rmd \bq \, .
\end{equation}
In order to make these functions optimally concentrated within $\mathcal{R}$, the following ratio of the energy contained within $\mathcal{R}$ as compared to the total energy in $\mathbb{R}^2$ may be defined
\begin{eqnarray}
    \lambda &=& \frac{\int_{\mathcal{R}} g^2(\bx) \rmd \bx}{\int_{\mathbb{R}^2} g^2(\bx) \rmd \bx} \, \label{eq:lambda_ratio}\\
    &=& \frac{\int_{\mathcal{Q}}\int_{\mathcal{Q}} G^*(\bq) \, D(\bq, \bq') \, G(\bq) \, \rmd \bq \, \rmd\bq'}{\int_{\mathcal{Q}} |G(\bq)|^2 \, \rmd \bq} \, ,
\end{eqnarray}
where $D(\bq,\bq') = (2 \pi)^{-2} \int_{\mathcal{R}} e^{i (\bq' - \bq) \cdot \bx} \rmd \bx$ is hermitian. Maximizing the energy ratio $\lambda$, results in the spectral-domain Fredholm integral equation
\begin{equation}
    \int_{\mathcal{Q}} D(\bq, \bq') \, G(\bq') \, \rmd \bq' = \lambda \, G(\bq) \, , \qquad \bq \in \mathcal{Q}.
\end{equation}
This leads to an eigenvalue problem resulting in quantization to produce the subscript label $\alpha$. Depending on the extent of the contour in spectral space $\mathcal{Q}$, a total of N eigenfunctions would span the basis. Naturally, N would increase as larger and larger wavenumbers are included inside $\mathcal{Q}$, meaning finer and finer structures are intended to be resolved. By definition, $\alpha$ is arranged in the decreasing order of eigenvalues $1 \geq \lambda_1 \geq \lambda_2 \geq ... \geq \lambda_N$. As seen in Eqn.~(\ref{eq:lambda_ratio}), $\lambda$ indicates the degree of energy concentration of the eigenfunction within $\mathcal{R}$. Therefore, the eigenfunction maximally concentrated within $\mathcal{R}$ takes the label ``1", the second most concentrated eigenfunction takes the label ``2", and so on. By construction, $g(\bx)$ can be chosen to be orthonormal over $\mathbb{R}^2$ which would render them orthogonal over $\mathcal{R}$.
\begin{equation} \label{eq: ortho_slepian}
    \int_{\mathbb{R}^2} g_{\alpha}(\bx)  \, g_{\beta}(\bx) \, \rmd \bx = \delta_{\alpha \beta}\, , \qquad \int_{\mathcal{R}} g_{\alpha}(\bx)  \, g_{\beta}(\bx) \, \rmd \bx = \lambda_{\alpha} \, \delta_{\alpha \beta} \, .
\end{equation}
Finally, as SW11 calls it, the ``planar Shannon number" N2D is defined as the sum of all the eigenvalues $\lambda_{\alpha}$
\begin{equation} \label{eq: shannon_number_eq}
    \mathrm{N2D} = \sum_{\alpha=1}^{N} \lambda_{\alpha} \,,
\end{equation}
which indicates the total number of eigenfunctions which are optimally concentrated within $\mathcal{R}$ and are strictly \mbox{bandlimited} to within $\mathcal{Q}$. For Fig.~\ref{fig:shannon_number_fig}, the Shannon number N2D $\approx 27$. This is calculated by summing over all the $\lambda_{\alpha}$ and rounding off the result to the next higher integer. This is why the first 27 basis functions are used in decomposing and reconstructing the map of $\bnabla \cdot \bfv_h$.

\bibliography{references}{}
\bibliographystyle{aasjournal}

\end{document}

%% file: newcommands.tex
\def\pointM{{\mathscr{M}}}
\def\pointS{{\mathscr{S}}}

\def\scrL{{\mathscr{L}}}
\def\bfscrE{{\mathbf{\mathscr{E}}}}

\def\sfzero{{\mathsf{0}}}
\def\bfzero{{\mathbf{0}}}
\def\sfnabla{{\mathsf{\nabla}}}
\def\scrL{{\mathscr{L}}}
\def\scrN{{\mathscr{N}}}
\def\scrS{{\mathscr{S}}}
\def\scrD{{\mathscr{D}}}
\newcommand{\bfomega}{\mbox{\boldmath $\bf \omega$}}
\newcommand{\bfsigma}{\mbox{\boldmath $\bf \sigma$}}
\newcommand{\bftau}{\mbox{\boldmath $\bf \tau$}}
\newcommand{\bfeta}{\mbox{\boldmath $\bf \eta$}}
\newcommand{\bftheta}{\mbox{\boldmath $\bf \theta$}}
\newcommand{\bflambda}{\mbox{\boldmath $\bf \lambda$}}
\newcommand{\bfphi}{\mbox{\boldmath $\bf \phi$}}
\newcommand{\bfpsi}{\mbox{\boldmath $\bf \psi$}}
\newcommand{\bfxi}{\mbox{\boldmath $\bf \xi$}}
\newcommand{\bfnu}{\mbox{\boldmath $\bf \nu$}}
\newcommand{\bfepsilon}{\mbox{\boldmath $\bf \epsilon$}}
\newcommand{\bfGamma}{\mbox{\boldmath $\bf \Gamma$}}
\newcommand{\bfvarepsilon}{\mbox{\boldmath $\bf \varepsilon$}}
\newcommand{\sftau}{\mbox{${\sf \tau}$}}
\newcommand{\sfLambda}{\mbox{${\sf \Lambda}$}}
\newcommand{\sfDelta}{\mbox{${\sf \Delta}$}}
\newcommand{\sfOmega}{\mbox{${\sf \Omega}$}}
\newcommand{\sfGamma}{\mbox{${\sf \Gamma}$}}
\newcommand{\sfQ}{\mbox{${\sf Q}$}}
\newcommand{\sfZ}{\mbox{${\sf Z}$}}
\newcommand{\sfV}{\mbox{${\sf V}$}}
\newcommand{\sfH}{\mbox{${\sf H}$}}
\newcommand{\sfX}{\mbox{${\sf X}$}}
\newcommand{\sfD}{\mbox{${\sf D}$}}
\newcommand{\bfOmega}{\mbox{\boldmath $\bf \Omega$}}
\newcommand{\bfPhi}{\mbox{\boldmath $\bf \Phi$}}
\newcommand{\bfPsi}{\mbox{\boldmath $\bf \Psi$}}
\newcommand{\bfLambda}{\mbox{\boldmath $\bf \Lambda$}}
\newcommand{\sfPhi}{\mbox{${\sf \Phi}$}}
\newcommand{\bfnabla}{\mbox{\boldmath $\bf \nabla$}}
\newcommand{\bfXi}{\mbox{\boldmath $\bf \Xi$}}
\newcommand{\bfgamma}{\mbox{\boldmath $\bf \gamma$}}

\def\rmd{{\mathrm{d}}}
\def\rmD{{\mathrm{D}}}

\newcommand{\tdot}{\,.\hspace{-0.98 mm}\raise.6ex\hbox{.}
                   \hspace{-0.98 mm}\raise1.2ex\hbox{.}\,}

\def\sfa{{\textbf{\textsf{a}}}}
\def\sfb{{\textbf{\textsf{b}}}}
\def\sfc{{\textbf{\textsf{c}}}}
\def\sfd{{\textbf{\textsf{d}}}}
\def\sfe{{\textbf{\textsf{e}}}}
\def\sff{{\textbf{\textsf{f}}}}
\def\sfg{{\textbf{\textsf{g}}}}
\def\sfh{{\textbf{\textsf{h}}}}
\def\sfl{{\textbf{\textsf{l}}}}
\def\sfi{{\textbf{\textsf{i}}}}
\def\sfI{{\textbf{\textsf{I}}}}
\def\sfj{{\textbf{\textsf{j}}}}
\def\sfk{{\textbf{\textsf{k}}}}
\def\sfm{{\textbf{\textsf{m}}}}
\def\sfn{{\textbf{\textsf{n}}}}
\def\sfp{{\textbf{\textsf{p}}}}
\def\sfq{{\textbf{\textsf{q}}}}
\def\sfr{{\textbf{\textsf{r}}}}
\def\sfs{{\textbf{\textsf{s}}}}
\def\sft{{\textbf{\textsf{t}}}}
\def\sfu{{\textbf{\textsf{u}}}}
\def\sfv{{\textbf{\textsf{v}}}}
\def\sfw{{\textbf{\textsf{w}}}}
\def\sfx{{\textbf{\textsf{x}}}}
\def\sfy{{\textbf{\textsf{y}}}}
\def\sfz{{\textbf{\textsf{z}}}}

\def\bfa{{\mathbf{a}}}
\def\bfb{{\mathbf{b}}}
\def\bfc{{\mathbf{c}}}
\def\bfd{{\mathbf{d}}}
\def\bfe{{\mathbf{e}}}
\def\bff{{\mathbf{f}}}
\def\bfg{{\mathbf{g}}}
\def\bfh{{\mathbf{h}}}
\def\bfi{{\mathbf{i}}}
\def\bfj{{\mathbf{j}}}
\def\bfk{{\mathbf{k}}}
\def\bfl{{\mathbf{l}}}
\def\bfm{{\mathbf{m}}}
\def\bfn{{\mathbf{n}}}
\def\bfp{{\mathbf{p}}}
\def\bfq{{\mathbf{q}}}
\def\bfr{{\mathbf{r}}}
\def\bfs{{\mathbf{s}}}
\def\bft{{\mathbf{t}}}
\def\bfu{{\mathbf{u}}}
\def\bfv{{\mathbf{v}}}
\def\bfw{{\mathbf{w}}}
\def\bfx{{\mathbf{x}}}
\def\bfy{{\mathbf{y}}}
\def\bfz{{\mathbf{z}}}

\def\bfA{{\mathbf{A}}}
\def\bfB{{\mathbf{B}}}
\def\bfC{{\mathbf{C}}}
\def\bfD{{\mathbf{D}}}
\def\bfE{{\mathbf{E}}}
\def\bfF{{\mathbf{F}}}
\def\bfG{{\mathbf{G}}}
\def\bfH{{\mathbf{H}}}
\def\bfI{{\mathbf{I}}}
\def\bfJ{{\mathbf{J}}}
\def\bfK{{\mathbf{K}}}
\def\bfL{{\mathbf{L}}}
\def\bfM{{\mathbf{M}}}
\def\bfN{{\mathbf{N}}}
\def\bfP{{\mathbf{P}}}
\def\bfQ{{\mathbf{Q}}}
\def\bfS{{\mathbf{S}}}
\def\bfT{{\mathbf{T}}}
\def\bfU{{\mathbf{U}}}
\def\bfV{{\mathbf{V}}}
\def\bfX{{\mathbf{X}}}
\def\bfZ{{\mathbf{Z}}}

\newcommand{\tr}[1]{\textcolor[rgb]{1,0,0}{#1}}
\newcommand{\tb}[1]{\textcolor[rgb]{0,0,1}{#1}}
\newcommand{\tg}[1]{\textcolor[rgb]{0,1,0}{#1}}
\newcommand{\oneby}[1]{\frac{1}{#1}}
\newcommand{\pdif}[2]{\frac{\partial #1}{\partial #2}}
\newcommand{\pdiff}[2]{\frac{\partial^2 #1}{\partial {#2}^2}}
\newcommand{\pdiffxy}[3]{\frac{\partial^2 #1}{\partial #2 \partial #3}}
\renewcommand{\vec}[1]{\boldsymbol{#1}}

\def\bq{{\boldsymbol{q}}}
\def\bk{{\boldsymbol{k}}}
\def\bx{{\boldsymbol{x}}}
\def\by{{\boldsymbol{y}}}
\def\bz{{\boldsymbol{z}}}